\documentclass[%
prd,
nofootinbib,
superscriptaddress,
preprintnumbers
]{revtex4}
\usepackage{amsmath,amssymb,bm,float,mathrsfs}
\usepackage{graphicx}
\usepackage{dcolumn}
\usepackage{color}
\usepackage{hyperref}
\usepackage{comment}

  \newcommand{\beq}{\begin{equation}}
  \newcommand{\eeq}{\end{equation}}
  \newcommand{\al}[1]{\begin{align} #1 \end{align}}
  \newcommand{\bi}{\begin{itemize}}
  \newcommand{\ei}{\end{itemize}}
  \def\dd{\mathrm{d}}
  
  \def\pd{\partial}
  \newcommand{\ave}[1]{\left\langle #1 \right\rangle}

\begin{document}

\title{Skewness consistency relation in large-scale structure and test of gravity theory}


\author{Daisuke Yamauchi}
\email[Email: ]{yamauchi``at''jindai.jp}
\affiliation{
Faculty of Engineering, Kanagawa University, Kanagawa, 221-8686, Japan
}

\author{Shoya Ishimaru}
\affiliation{
Graduate School of Science and Engineering, Saga University, Saga 840-8502, Japan
}

\author{Takahiko Matsubara}
\email[Email: ]{tmats``at''post.kek.jp}
\affiliation{
Institute of Particle and Nuclear Studies, High Energy Accelerator Research Organization (KEK), Oho 1-1, Tsukuba 305-0801, Japan
}
\affiliation{
The Graduate University for Advanced Studies (SOKENDAI), Tsukuba, Ibaraki 305-0801, Japan
}

\author{Tomo Takahashi}
\email[Email: ]{tomot``at"cc.saga-u.ac.jp}
\affiliation{
Department of Physics, Saga University, Saga 840-8502, Japan
}

\begin{abstract}
We investigate the skewness of galaxy number density 
fluctuations
as a possible probe to test gravity theories.
We find that the specific linear combination of the skewness parameters corresponds to the coefficients of
the second-order kernels of the density contrast, 
which can be regarded as the consistency relation and used as a test of  
general relativity and modified gravity theories.
We also extend the analysis of the skewness parameters
from real space to redshift space and derive
the redshift-space skewness consistency relation.
\end{abstract}

\preprint{KEK-TH-2483, KEK-Cosmo-0305}

\maketitle

\section{Introduction}

It is one of the biggest challenges of modern cosmology to understand the physical origin 
of the present cosmic acceleration of the Universe.
It might eventually require the presence of a new type of energy component, usually called dark energy.
Another possibility
is that the accelerated cosmic expansion
could arise due to a modification of general
relativity on cosmological scales.
Among various cosmological observational data, measuring the growth history of density contrast of
large-scale structure is one of the most popular tools to test the nature of dark energy and the modification of
the theory of gravity responsible for the present cosmic acceleration.
Although current observational data 
have already constrained
the growth of structure at the linear level,
there is no evidence of the deviation from the value predicted by the standard $\Lambda$ cold dark matter 
($\Lambda$CDM) model.
Hence, further investigation of the cosmological model landscape requires introducing other observables
to capture the modification of gravity theory.

The power spectrum, or its Fourier counterpart, the two-point correlation function, is one of the most popular statistics
to characterize cosmological observables such as cosmic microwave background and large-scale structure~(e.g., \cite{Peebles book}).
Since this can fully characterize the random Gaussian field, it can be treated as a good statistical quantity
for large-scale structure on sufficiently large scales, or sufficiently early time, and cosmic microwave background.
The cosmological structure generally grows nonlinearly in time due to gravitational interaction, which naturally
introduces nonvanishing non-Gaussianity on small scale, or in late time, even in the case of the initial Gaussian field. 
Much information in such a non-Gaussian random field cannot be captured by solely using the power spectrum.
A straightforward way to extract information on the non-Gaussianity of such a field is to consider higher-order
spectra such as bispectrum, trispectrum, and so on (e.g., \cite{Bernardeau:2001qr}).
However, it is challenging to measure these higher-order correlation functions accurately, since these are 
functions of scales with many arguments.
Hence, it would be important to use other statistical tools for probing non-Gaussianities of random fields.
The skewness, which is a measure of the asymmetry of the probability distribution of a random variable about its mean,
is one of the popular measures to investigate the non-Gaussianity.
The relation between higher-order spectra and the skewness parameters is derived analytically when
the non-Gaussianity is weak~\cite{Matsubara:1994wn,Matsubara:2003yt}.
In particular, the leading order contributions of the higher-order spectra to the skewness parameters are determined
through the integration of wavevectors of the bispectrum.
Therefore,
the skewness parameters can be used to extract the information of 
the bispectrum and higher-order spectra for the nonlinear density field.

In this paper, we consider the quasi-nonlinear growth of large-scale structure as a way to provide
new insight into the modification of the gravity theory that would not be imprinted in the linear perturbation theory.
It has been discussed in the literature that such a quasi-nonlinearity can be explored by observing the higher-order spectra of galaxies~\cite{Takushima:2013foa,Takushima:2015iha,Yamauchi:2017ibz,Hirano:2018uar,Crisostomi:2019vhj,Lewandowski:2019txi,Hirano:2020dom,Yamauchi:2021nxw} and cosmic microwave background~\cite{Namikawa:2018erh}.
In particular, a possible parameter to trace the nonlinear growth of structure is 
the time evolution of the coefficients of the second-order kernels for density contrast, the so-called second-order indices,
which is originally developed in Ref.~\cite{Yamauchi:2017ibz,Namikawa:2018erh}.
However, Ref.~\cite{Yamauchi:2021nxw} pointed out that the features of the quasi-nonlinear growth are partially hidden
by the uncertainty of the nonlinear galaxy bias functions.
In order to extract the information of the modification of gravity theory obtained from the quasi-nonlinear growth,
we need to carefully construct the combination of observables.
In this paper, we focus on the skewness of the galaxy number density fluctuations
as another probe of
the underlying gravity theory.
In particular, we use not only a common skewness parameter of the density field
but also other skewness parameters which involve spatial derivatives of the density field.
We investigate how the effects of the quasi-nonlinear growth appear in the skewness parameters and 
construct their combination of the skewness parameters that can be used to extract the information of
the modification of gravity theory.
Even in the case of the $\Lambda$CDM Universe with general relativity, the resultant expression
represents the consistency relation between the skewness parameters.

This paper is organized as follows. In Sec.\ref{sec:Consistency relation in real space}, we summarize the skewness parameters
derived in the previous works and develop the skewness parameters for the galaxy number density fluctuations as one of the actual observables
of large-scale structure observations.
We then derive the single consistency relation for the three real-space skewness parameters.
In Sec.~\ref{sec:Consistency relation in redshift space}, we extend the analysis to the redshift space, in which the effect of
the anisotropic effect due to the redshift-space distortion should be properly included.
We then derive the three consistency relations for the five independent redshift-space skewness parameters.
Finally, Sec.~\ref{sec:Summary} is devoted to a summary and conclusion.
Throughout this paper, as our fiducial model, we assume a $\Lambda$CDM cosmological model with parameters:
$\Omega_{{\rm m},0}=0.3111$, $\Omega_{{\rm b},0}=0.0490$, $\Omega_{\Lambda}=0.6889$, 
$h=0.6766$, $A_{\rm s}=2.105\times 10^{-9}$, 
$n_{\rm s}=0.9665$, and $k_{\rm pivot}=0.05\,[{\rm Mpc}^{-1}]$.

\section{Skewness consistency relation in real space}
\label{sec:Consistency relation in real space}

In this section, we first summarize 
the description of
the skewness parameters, following \cite{Matsubara:2003yt,Matsubara:2020knr}.
We then calculate 
the skewness parameters from
the bispectrum for the galaxy number density 
fluctuations, 
which includes the effects of
the nonlinear gravitational clustering and galaxy biasing, as natural observables of large-scale structure observations. 
Combining the resultant expressions,
we then derive a consistency relation of skewness parameters in real space.

\subsection{Real-space skewness parameters}

We consider three-dimensional density field $\rho ({\bm x})$. We denote the density contrast
by $\delta ({\bm x})=\rho ({\bm x})/\ave{\rho ({\bm x})}-1$, where $\ave{\cdots}$ represents the ensemble average.
We assume that the density contrast is smoothed by a smoothing function $W_R$ with smoothing length
$R$, defined as
\al{
	\delta_R ({\bm x})=\int\dd^3{\bm x}^\prime W_R(|{\bm x}-{\bm x}^\prime |)\delta ({\bm x}^\prime )
	\,.
}
In this paper, we use a Gaussian kernel function to obtain the smooth density field as 
\al{
	W_R(x)=\frac{e^{-x^2/(2R^2)}}{(2\pi)^{3/2}R^3}
	\,.
}
The variance
of density contrast smoothed on a scale $R$ is defined as
\al{
	\sigma_0^2=\ave{\delta_R^2}
	\,.
}
Let us introduce the three real-space skewness parameters, which are defined as
\al{
&S_{\rm r}^{(0)}=\frac{\ave{\delta_R^3}_{\rm c}}{\sigma_0^4}
	\,,\label{eq:S0}\\
	&S_{\rm r}^{(1)}=\frac{3}{2}\frac{\ave{\delta_R|\nabla\delta_R|^2}_{\rm c}}{\sigma_0^2\sigma_1^2}
	\,,\\
	&S_{\rm r}^{(2)}=-\frac{9}{4}\frac{\ave{|\nabla\delta_R|^2\Delta\delta_R}_{\rm c}}{\sigma_1^4}
	\,,\label{eq:S2}
}
where $\ave{\cdots}_{\rm c}$ denotes the three-point cumulants and we have defined $\sigma_1$ as 
\al{
	\sigma_1^2=\ave{|\nabla\delta_R|^2}
	\,.
}

To connect the statistical properties of the density contrast with the skewness parameters, 
it is convenient to consider the Fourier transform of the density field as
\al{
	\widetilde\delta ({\bm k})=\int\dd^3{\bm x}e^{-i{\bm k}\cdot{\bm x}}\delta ({\bm x})
	\,.
}
The Fourier transform of the smoothed density contrast is given by 
$\widetilde\delta_R({\bm k})=\widetilde\delta ({\bm k})W(kR)$\,,
where $W(kR)$ denotes the Fourier transform of the kernel function $W_R(x)$ as
$W(kR)=\int\dd^3{\bm x}e^{-i{\bm k}\cdot{\bm x}}W_R(x)$. For the Gaussian smoothing, we have $W(kR)=e^{-k^2R^2/2}$.
The power spectrum and bispectrum of the density contrast 
in the Fourier space have the forms
\al{
	&\ave{\widetilde\delta ({\bm k}_1)\widetilde\delta ({\bm k}_2)}
		=(2\pi)^3\delta_{\rm D}^3({\bm k}_{12})P(k_1)
	\,,\\
	&\ave{\widetilde\delta ({\bm k}_1)\widetilde\delta ({\bm k}_2)\widetilde\delta ({\bm k}_3)}
		=(2\pi)^3\delta_{\rm D}^3({\bm k}_{123})B(k_1,k_2,k_3)
	\,,
}
where ${\bm k}_{ij}={\bm k}_i+{\bm k}_j$, ${\bm k}_{ijk}={\bm k}_i+{\bm k}_j+{\bm k}_k$\,.
With these notations, the skewness parameters are 
written as~\cite{Matsubara:2020knr}
\al{
	S_{\rm r}^{(a)}=\frac{1}{\sigma_0^{4-2a}\sigma_1^{2a}}
			\int\frac{\dd^3{\bm k}_1}{(2\pi)^3}\frac{\dd^3{\bm k}_2}{(2\pi)^3}\frac{\dd^3{\bm k}_3}{(2\pi)^3}
			(2\pi)^3\delta_{\rm D}^3({\bm k}_{123})s^{(a)}({\bm k}_1,{\bm k}_2,{\bm k}_3)B(k_1,k_2,k_3)W(k_1R)W(k_2R)W(k_3R)
	\,,\label{eq:real-space skew}
}
where the kernel functions are given by
\al{
	&s^{(0)}=1
	\,,\\
	&s^{(1)}=-\frac{3}{2}{\bm k}_1\cdot{\bm k}_2
	\,,\\
	&s^{(2)}=-\frac{9}{4}({\bm k}_1\cdot{\bm k}_2)k_3^2
	\,.
}
Thus, once the bispectrum for the density contrast is specified, we can calculate the skewness parameters by using the formula Eq.~\eqref{eq:real-space skew}.

\subsection{Reduced formula for real-space galaxy bispectrum}
\label{sec:Reduced formula for real-space galaxy bispectrum}

In the standard cosmological perturbation theory, gravitational nonlinear growth of structure gives non-negligible
contributions to the nonlinear density field, which naturally leads to the non-vanishing bispectrum for the density field.
Moreover, the density contrast is indirectly related to the observables of large-scale structure such as the galaxy number density fluctuations.
Hence, it is natural to apply the formulation derived in the previous subsection to the bispectrum of the galaxy number density fluctuations, which
includes the effects of the nonlinear gravitational clustering and the galaxy biasing.
In the standard perturbation theory, the nonlinear density contrast in Fourier space, $\widetilde\delta({\bm k})$,
is formally expanded in terms of the linear density contrast $\widetilde\delta_{\rm L}({\bm k})$ as
\al{
	\widetilde\delta({\bm k})
		=\widetilde\delta_{\rm L}({\bm k})
			+\int\frac{\dd^3{\bm k}_1}{(2\pi)^3}\frac{\dd^3{\bm k}_2}{(2\pi)^3}
			(2\pi)^3\delta_{\rm D}^3({\bm k}_{12}-{\bm k})F_2({\bm k}_1,{\bm k}_2)
			\widetilde\delta_{\rm L}({\bm k}_1)\widetilde\delta_{\rm L}({\bm k}_2)+\cdots
	\,.\label{eq:delta exp}
}
Here, the explicit form of the kernel function $F_2$ will be shown in the subsequent subsection.
To connect the nonlinear density field $\delta$ with the galaxy number density fluctuations $\delta_{\rm g}$, we need some relation.
We assume that the galaxy number density fluctuations
$\delta_{\rm g}({\bm x})$ up to the second-order in the perturbative expansion
of the density contrast $\delta({\bm x})$
is well described by the combination of the linear bias $b_1$, the second-order bias $b_2$, and the tidal bias $b_{K^2}$ as \cite{Desjacques:2016bnm}
\al{
	\delta_{\rm g}({\bm x})=b_1\delta ({\bm x})
            +\frac12 b_2\Bigl\{[\delta({\bm x})]^2-\ave{\delta^2}\Bigr\}
			+b_{K^2}\Bigl\{[s_{ij}({\bm x})]^2-\ave{s_{ij}^2}\Bigr\}
			+\cdots
	\,,
}
where
\al{
    s_{ij}=\left( \frac{\pd_i \pd_j}{\pd^2} - \frac13 \delta_{ij} 
			\right) \delta (\bm x)
   \,.
}
Here, the terms $\ave{\delta^2}$ and $\ave{s_{ij}^2}$ ensure the condition $\ave{\delta_{\rm g}({\bm x})}=0$.
Combining these, the Fourier transform of the galaxy number density fluctuations
$\widetilde\delta_{\rm g}({\bm k})$ in real space 
can be written in terms of the linear density contrast $\widetilde\delta_{\rm L}({\bm k})$:
\al{
	\widetilde\delta_{\rm g}({\bm k})
		=b_1\widetilde\delta_{\rm L}({\bm k})
			+\int\frac{\dd^3{\bm k}_1}{(2\pi)^3}\frac{\dd^3{\bm k}_2}{(2\pi)^3}(2\pi)^3\delta_{\rm D}^3({\bm k}_{12}-{\bm k})
				Z_{2,{\rm r}}({\bm k}_1,{\bm k}_2)\widetilde\delta_{\rm L}({\bm k}_1)\widetilde\delta_{\rm L}({\bm k}_2)
	+\cdots
	\,.
}
where
\al{
	Z_{2,{\rm r}}({\bm k}_1,{\bm k}_2)
		:=b_1F_2({\bm k}_1,{\bm k}_2)+\frac{1}{2}b_2
					+b_{K^2}\left\{(\widehat{\bm k}_1\cdot\widehat{\bm k}_2)^2-\frac{1}{3}\right\}
	\,,\label{eq:Z2 real}
}
with $\widehat{\bm k}_a:={\bm k}_a/|{\bm k}_a|$.
We now assume that the linear density contrast $\widetilde\delta_{\rm L}({\bm k})$ obeys the Gaussian statistics with the power spectrum
$P_{\rm L}(k)$ defined through  $\ave{\widetilde\delta_{\rm L}({\bm k}_1)\widetilde\delta_{\rm L}({\bm k}_2)}=(2\pi)^3\delta_{\rm D}^3({\bm k}_{12})P(k_1)$.
Hereafter, 
we consider the skewness parameters for the galaxy number density fluctuations instead of the density contrast
since $\delta_g$ is the actual observable obtained 
from large-scale structure observations.

The power spectrum and bispectrum of galaxy number density fluctuations
are defined as
\al{
    &\ave{\widetilde\delta_{\rm g}({\bm k}_1)\widetilde\delta_{\rm g}({\bm k}_2)}
        =(2\pi)^3\delta_{\rm D}^3({\bm k}_{12})P_{\rm g}(k_1)
    \,,\\
    &\ave{\widetilde\delta_{\rm g}({\bm k}_1)\widetilde\delta_{\rm g}({\bm k}_2)\widetilde\delta_{\rm g}({\bm k}_3)}
        =(2\pi)^3\delta_{\rm D}^3({\bm k}_{12})B_{\rm g}(k_1,k_2,k_3)
    \,.
}
In the lowest-order approximation of the perturbation theory, the real-space galaxy bispectrum due to the nonlinear growth
is given by
\al{
	B_{\rm g,r}(k_1,k_2,k_3)=2Z_{2,{\rm r}}({\bm k}_1,{\bm k}_2)b_1^2P_{\rm L}(k_1)P_{\rm L}(k_2)+\text{(cyc.)}
	\,,\label{eq:real-space galaxy bispectrum}
}
where ``$+({\rm cyc.})$'' denotes the cyclic permutation with respect to the arguments ${\bm k}_1$, ${\bm k}_2$, and ${\bm k}_3$.

Let us evaluate the skewness parameters with the Gaussian smoothing kernel, following the method 
originally developed in Ref.~\cite{Matsubara:1994wn}.
The galaxy skewness parameters we consider in this paper are given by
\al{
    &S_{\rm g,r}^{(0)}=\frac{\ave{\delta_{\rm g}^3}_{\rm c}}{\sigma_{{\rm g}0}^4}
	\,,\\
	&S_{\rm g,r}^{(1)}=\frac{3}{2}\frac{\ave{\delta_{\rm g}|\nabla\delta_{\rm g}|^2}_{\rm c}}{\sigma_{{\rm g}0}^2\sigma_{{\rm g}1}^2}
	\,,\\
	&S_{\rm g,r}^{(2)}=-\frac{9}{4}\frac{\ave{|\nabla\delta_{\rm g}|^2\Delta\delta_{\rm g}}_{\rm c}}{\sigma_{{\rm g}1}^4}
	\,,
}
where the quantities
$\sigma_{{\rm g}j}$ for the galaxy number density 
fluctuations are given by
\al{
	\sigma_{{\rm g}j}^2=b_1^2\int\dd\ln k\,k^{2j}\frac{k^3}{2\pi^2}P_{\rm L}(k)W^2(kR)
	\,.
}
Using the expression of the galaxy bispectrum Eq.~\eqref{eq:real-space galaxy bispectrum}, we obtain the real-space galaxy skewness parameters as
\al{
	S_{\rm g,r}^{(a)}=\frac{b_1^2}{\sigma_{{\rm g}0}^{4-2a}\sigma_{{\rm g}1}^{2a}}
			\int\frac{\dd^3{\bm k}_1}{(2\pi)^3}\frac{\dd^3{\bm k}_2}{(2\pi)^3}
			s^{(a)}({\bm k}_1,{\bm k}_2)\bigg[2Z_{2,{\rm r}}({\bm k}_1,{\bm k}_2)P_{\rm L}(k_1)P_{\rm L}(k_2)+({\rm cyc.})\biggr]
			e^{-(k_1^2+k_2^2+{\bm k}_1\cdot{\bm k}_2)R^2}
	\,.
}
To evaluate the skewness parameters, it would be convenient to introduce 
the symmetrized kernel function $s^{(a),{\rm sym}}$
defined as
\al{
	s^{(a),{\rm sym}}({\bm k}_1,{\bm k}_2,{\bm k}_3)
		:=\frac{1}{3}s^{(a)}({\bm k}_1,{\bm k}_2,{\bm k}_3)+(\text{cyc.})
	\,,
}
where $({\rm cyc.})$ denotes the cyclic permutations of the previous term.
The explicit calculation shows
\al{
	&s^{(0),{\rm sym}}=1
	\,,\\
	&s^{(1),{\rm sym}}=\frac{1}{2}\left( k_1^2+k_2^2+{\bm k}_1\cdot{\bm k}_2\right)
	\,,\label{eq:s^1 def}\\
	&s^{(2),{\rm sym}}=\frac{3}{2}\Bigl[ k_1^2k_2^2-({\bm k}_1\cdot{\bm k}_2)^2\Bigr]
	\,.\label{eq:s^2 def}
}
Once we replace $s^{(a)}\to s^{(a),{\rm sym}}$ in Eq.~\eqref{eq:real-space skew}, the bispectrum for
the density contrast can be replaced 
by the following function:
\al{
	S_{\rm g,r}^{(a)}=\frac{b_1^2}{\sigma_{{\rm g}0}^{4-2a}\sigma_{{\rm g}1}^{2a}}
			\int\frac{\dd^3{\bm k}_1}{(2\pi)^3}\frac{\dd^3{\bm k}_2}{(2\pi)^3}
			s^{(a),{\rm sym}}({\bm k}_1,{\bm k}_2)\, 6Z_{2,{\rm r}}({\bm k}_1,{\bm k}_2)P_{\rm L}(k_1)P_{\rm L}(k_2)
			e^{-(k_1^2+k_2^2+{\bm k}_1\cdot{\bm k}_2)R^2}
	\,.
}
By using the rotational invariance, we can reduce the dimensionality of the integrals for the skewness parameters.
Introducing the coordinate system as
\al{
	&{\bm k}_1=\frac{p}{R}\left(\sqrt{1-\nu^2},0,\nu\right)
	\,,\ \ \ 
	{\bm k}_2=\frac{q}{R}\left( 0,0,1\right)
	\,,\label{eq:coordinate real space}
}
the integrals for the skewness parameters 
analytically reduce to
\al{
	S_{\rm g,r}^{(a)}=\frac{b_1^3}{8\pi^4R^{2a+6}\sigma_{{\rm g}0}^{4-2a}\sigma_{{\rm g}1}^{2a}}
			\int_0^\infty\dd p\int_0^\infty\dd q
			e^{-(p^2+q^2)}	\widetilde S_{\rm g,r}^{(a)}(p,q)
			P_{\rm L}\left(\frac{p}{R}\right)P_{\rm L}\left(\frac{q}{R}\right)
	\,,
}
where $\widetilde S_{\rm g,r}^{(a)}(p,q)$ are kernel functions defined as
\al{
	\widetilde S_{\rm g,r}^{(a)}(p,q)
		=\frac{6p^2q^2}{b_1}\int_{-1}^1\dd\nu
			\widetilde s^{(a),{\rm sym}}(p,q,\nu)Z_{2,{\rm r}}(p,q,\nu )e^{-pq\nu}
	\,,\label{eq:tilde S^a}
}
with
\al{
	&\widetilde s^{(0),{\rm sym}}=1
	\,,\label{eq:s^0}\\
	&\widetilde s^{(1),{\rm sym}}=\frac{1}{2}\left(p^2+q^2+pq\nu\right)
	\,\label{eq:s^1},\\
	&\widetilde s^{(2),{\rm sym}}=\frac{3p^2q^2}{2}\left(1-\nu^2\right)
	\,.\label{eq:s^2}
}
Therefore, once the explicit expression of the second-order kernel of the density contrast, $F_2$, is given, we evaluate
the skewness parameters by integrating the above formula.

\subsection{Consistency relation in real space}

To proceed the analysis, let us consider the specific type of the second-order kernel of the density contrast, $F_2$.
In this paper, we will consider the following form~\cite{Takushima:2013foa,Takushima:2015iha,Yamauchi:2017ibz,Hirano:2018uar,Crisostomi:2019vhj,Lewandowski:2019txi,Hirano:2020dom,Yamauchi:2021nxw}:
\al{
	F_2({\bm k}_1,{\bm k}_2)
		=\kappa\biggl[ 1+\frac{1}{2}(\widehat{\bm k}_1\cdot\widehat{\bm k}_2)\left(\frac{k_1}{k_2}+\frac{k_2}{k_1}\right)\biggr]
			-\frac{2}{7}\lambda\biggl[1-(\widehat{\bm k}_1\cdot\widehat{\bm k}_2)^2\biggr]
	\,,\label{eq:F2}
}
where $\kappa$ and $\lambda$ are time-dependent parameters corresponding to the coefficients for the terms with characteristic 
wavevector dependence and they depend on the underlying gravity theory.
It is known that this specific form of the second-order kernels describes a wide variety of the modified theory of gravity
since the modification of gravity theory alters the clustering properties of nonlinear structures.
In the Einstein-de Sitter Universe with general relativity, one finds $\kappa=\lambda =1$~\cite{Takushima:2013foa}, 
while in the case of the $\Lambda$CDM Universe,
$\lambda$ slightly deviates from unity. 
In the limit of $\Omega_{\rm m}\approx 1$, $\lambda$ can be well approximated as 
$\lambda\approx 1-\frac{3}{572}(1-\Omega_{\rm m})$~\cite{Yamauchi:2017ibz,Bouchet:1994xp,Bernardeau:2001qr}.
If the gravity theory is described by the Horndeski scalar-tensor gravity theories~\cite{Horndeski:1974wa,Deffayet:2011gz,Kobayashi:2011nu},
$\kappa$ still takes the standard value, but the deviation of $\lambda$ 
from unity contains the information from the underlying gravity model~\cite{Takushima:2013foa,Yamauchi:2017ibz}. 
In the case of a broader class of scalar-tensor gravity theories called degenerate higher-order scalar-tensor (DHOST) theories~\cite{Langlois:2015cwa,Crisostomi:2016czh,Achour:2016rkg,BenAchour:2016fzp,Langlois:2017dyl}
(see \cite{Langlois:2018dxi,Kobayashi:2019hrl} for a review)
which is a natural extension of the Horndeski theory, 
both $\kappa$ and $\lambda$ can depend on time and deviate from unity~\cite{Hirano:2018uar,Crisostomi:2019vhj,Lewandowski:2019txi,Hirano:2020dom,Yamauchi:2021nxw}.
Therefore, the time dependence of the coefficients of the second-order kernel, namely $\kappa$ and $\lambda$, yields
a powerful probe of modified gravity theories. 
Hereafter, we consider how we can extract the information of $\kappa$ and $\lambda$ from  
observations of the skewness parameters.

In order to evaluate the skewness parameters in this setup, we substitute 
Eq.~\eqref{eq:F2} into Eq.~\eqref{eq:Z2 real} with the coordinate
of Eq.~\eqref{eq:coordinate real space} to obtain 
the real-space $Z_2$ kernel as
\al{
	\frac{2}{b_1}Z_{2,{\rm r}}(p,q,\nu )
		=g_0\mathsf{P}_0+g_1\mathsf{P}_1+g_2\mathsf{P}_2
	\,,
}
where 
\al{
    \mathsf{P}_0=P_0(\nu)
    \,,\ \ 
    \mathsf{P}_1=\left(\frac{p}{q}+\frac{q}{p}\right)P_1(\nu )
    \,,\ \ 
    \mathsf{P}_2=P_2(\nu )
    \,,
	\label{eq:mathsfP real def}
}
with $P_i(\nu )$ being the Legendre polynomials.
Here, we have introduced $g_i$ ($i=0,1,2$), which are written
in terms of $\kappa$, $\lambda$, and the bias functions as
\al{
	&g_0=2\kappa -\frac{8}{21}\lambda 
	+\frac{b_2}{b_1}
	\,,\label{eq:g0}\\
	&g_1=\kappa
	\,,\\
	&g_2=\frac{4}{3}\left(\frac{2}{7}\lambda+\frac{b_{K^2}}{b_1}\right)
	\,.\label{eq:g2}
}
In the case of the Einstein-de Sitter Universe without introducing the nonlinear bias factors, these 
are written as $g_0=34/21,g_1=1$, and $g_2=8/21$.
The kernel functions given in
Eq.~\eqref{eq:tilde S^a} reduce to
\al{
	&\widetilde S_{\rm g,r}^{(0)}
		=-6\Bigl[g_1(p^2+q^2)+3g_2\Bigr]\left(\cosh (pq)-\frac{\sinh(pq)}{pq}\right)+6(g_0+g_2)pq\sinh (pq)
	\,,\label{eq:tilde S^0}\\
	&\widetilde S^{(1)}_{\rm g,r}
		=-3\Bigl[g_0p^2q^2+g_1(p^2+q^2)(2+p^2+q^2)
  +g_2(3+p^2)(3+q^2)\Bigr]\left(\cosh(pq)-\frac{\sinh(pq)}{pq}\right)
	\notag\\
	&\qquad\qquad
			+3\Bigl[g_0(p^2+q^2)+g_1(p^2+q^2)+g_2(3+p^2+q^2)\Bigr]pq\sinh (pq)
	\,,\\
	&\widetilde S^{(2)}_{\rm g,r}
		=18\Bigl[ g_0p^2q^2+3g_1(p^2+q^2)+(18+p^2q^2)g_2\Bigr]\left(\cosh (pq)-\frac{\sinh (pq)}{pq}\right)
			-18\Bigl[ g_1(p^2+q^2)+6g_2\Bigr] pq\sinh (pq)
	\,.\label{eq:tilde S^2}
}
When substituting $g_0=34/21,g_1=1$, and $g_2=8/21$ into the above expressions, one can easily reproduce
the results derived in \cite{Matsubara:2020knr}.
An important observation is that all the parameters $g_i$ linearly appear in the above expression of the skewness parameters.
Therefore, $g_i$ can be rewritten in terms of the linear combination of the skewness parameters.
To do so, it would be convenient to introduce the matrix $\mathsf{M}_{\rm r}$, whose components are given by
\al{
	\mathsf{M}_{\rm r}^{(a)}{}_{(i)}:=\frac{\pd S_{\rm g,r}^{(a)}}{\pd g_i}=\frac{b_1^3}{8\pi^4R^{2a+6}\sigma_0^{4-2a}\sigma_1^{2a}}
			\int_0^\infty\dd p\int_0^\infty\dd q
			e^{-(p^2+q^2)}\widetilde{\mathsf{M}}_{\rm r}^{(a)}{}_{(i)}(p,q)
			P_{\rm L}\left(\frac{p}{R}\right)P_{\rm L}\left(\frac{q}{R}\right)
	\,.\label{eq:M^a_i}
}
Here one can easily extract the explicit form of $\widetilde{\mathsf{M}}_{\rm r}^{(a)}{}_{(i)}(p,q):=\pd\widetilde S_{\rm r}^{(a)}(p,q)/\pd g_i$ 
from Eqs.~\eqref{eq:tilde S^0}--\eqref{eq:tilde S^2}.
Once the shape of the linear power spectrum $P_{\rm L}(k)$ is given, the matrix components can be obtained 
by integrating Eq.~\eqref{eq:M^a_i} numerically.
With this convention, the set of 
three equations for the skewness parameters can be rewritten in a simpler form
\al{
	S_{\rm g,r}^{(a)}=\sum_{i=0}^2\mathsf{M}_{\rm r}^{(a)}{}_{(i)}\,g_i
	\,.
}
At least in the standard cosmology with our fiducial parameter set, it is numerically confirmed that the determinant of $\mathsf{M}$ is nonzero.
We therefore solve the above equation in terms of $g_i$ as
\al{
	g_i=\sum_{a=0}^2\,[\mathsf{M}_{\rm r}^{-1}]^{(i)}{}_{(a)}S_{\rm g,r}^{(a)}
	\,.\label{eq:consistency real space}
}
This is one of the main results of this paper.
However, $\kappa$ and $\lambda$ always appear with the nonlinear galaxy bias function
in $g_0$ and $g_2$ terms, as seen in Eqs.~\eqref{eq:g0} and \eqref{eq:g2}, 
which indicates that
there is a strong degeneracy between $\kappa$, $\lambda$, $b_2$, and $b_{K^2}$.
The signals of gravity theory existing in $g_0$ and $g_2$ terms would be hidden by
the uncertainty of the nonlinear bias factors.
Therefore, the meaningful consistency relation comes only from $g_1$ term:
\al{
	g_1=\kappa =[\mathsf{M}_{\rm r}^{-1}]^{(1)}{}_{(0)}S_{\rm g,r}^{(0)}+[\mathsf{M}_{\rm r}^{-1}]^{(1)}{}_{(1)}S_{\rm g,r}^{(1)}+[\mathsf{M}_{\rm r}^{-1}]^{(1)}{}_{(2)}S_{\rm g,r}^{(2)}
	\,.\label{eq:real space consistency}
}
As discussed above, the confirmation of the deviation of $g_1=\kappa$ from unity 
would indicate a departure from the Einstein gravity, in particular, the specific types of the modification of gravity theory,
the DHOST theory beyond-Horndeski class~\cite{Hirano:2018uar,Crisostomi:2019vhj,Lewandowski:2019txi,Hirano:2020dom,Yamauchi:2021nxw}, 
which can be tested once we measure the skewness parameters. On the other hand, 
when gravity is described by the Horndeski scalar-tensor theory including general relativity,
we have
$g_1=\kappa =1$, 
therefore Eq.~\eqref{eq:real space consistency} can be regarded as the consistency relation among the three skewness parameters in the theory. 

\begin{figure}
\includegraphics[width=100mm]{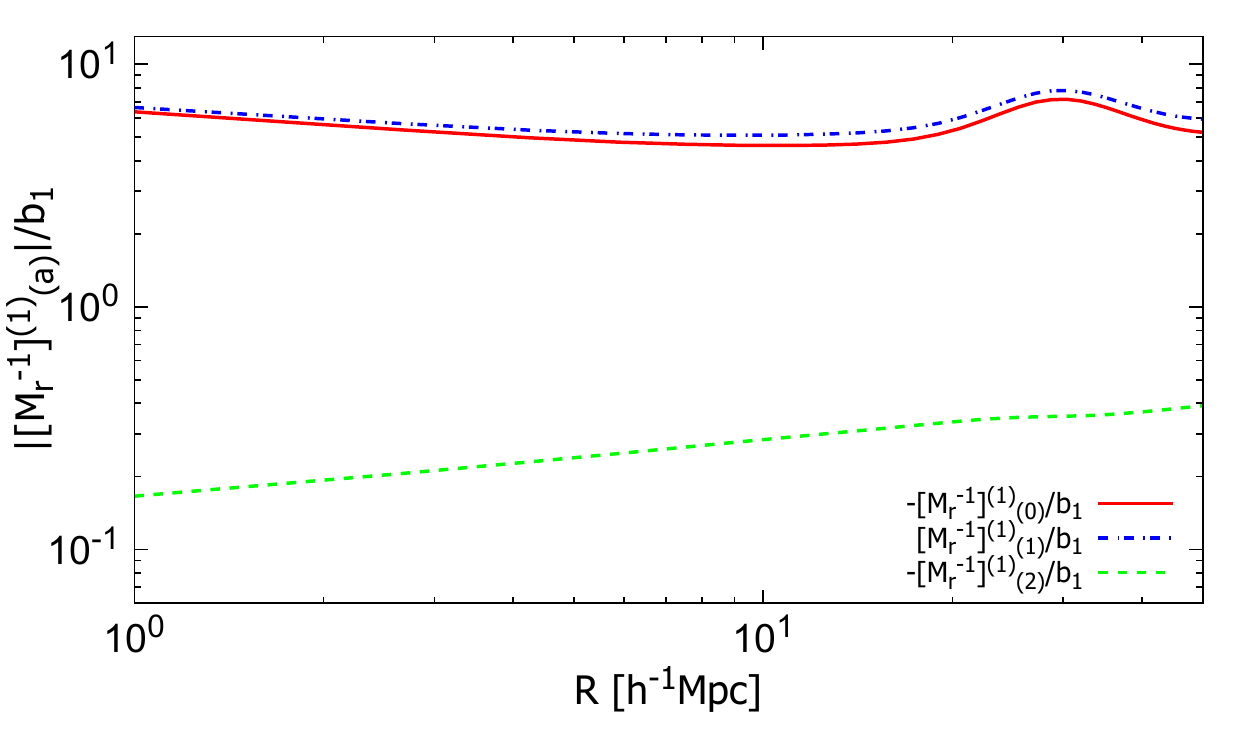}
\caption{
The absolute value of the coefficients appearing in the consistency relation for the real-space skewness parameters, $[\mathsf{M}_{\rm r}^{-1}]^{(1)}{}_{(a)}/b_1$,
as a function of the smoothing length $R\,[h^{-1}{\rm Mpc}]$ with
$a=0$ (red solid), $a=1$ (blue dot-dashed), and $a=2$ (green dashed).
}
\label{fig:MInv_real-space}
\end{figure} 

Before closing this section, as a demonstration, 
we show the consistency relation for
the real-space skewness parameters
Eq.~\eqref{eq:real space consistency} 
with explicit numerical coefficients for a given smoothing scale.
To calculate it explicitly, we need to specify
the statistical nature of the linear density field.
The linear density field $\delta_{\rm L}$ is related
to the primordial curvature perturbation $\Phi$ through
\al{
    \delta_{\rm L}({\bm k})
        ={\cal M}(k)\Phi ({\bm k})
    \,,
}
where the function ${\cal M}(k)$ is defined as
\al{
    {\cal M}(k,z)=\frac{2}{3}\frac{D_+(z)}{D_+(z_\ast)(1+z_\ast)}\frac{k^2T(k)}{H_0^2\Omega_{\rm m,0}}
    \,,
}
with $D_+(z)$ being the linear growth rate, and
$z_\ast$ represents an arbitrary redshift
at the matter-dominated era.
$T(k)$ is the matter transfer function normalized to
unity on large scales~\cite{Eisenstein:1997ik}.
The power spectrum for the linear
density field is related to that of
the primordial perturbations as
$P_{\rm L}(k)={\cal M}^2(k)P_\Phi (k)$,
where $P_\Phi (k)$ denotes the power spectrum
for the primordial curvature perturbations.
With these, we now calculate 
the matrix components of $\mathsf{M}_{\rm r}$ and 
the corresponding consistency relation:
\al{
	&\mathsf{M}_{\rm r}
		=\frac{1}{b_1}
        \left(
		\begin{array}{ccc}
		3.4930 & -2.1752 & 0.21206 \\
		3.4049 & -1.9486 & 0.17288 \\
		4.2484 & -3.0878 & -0.34998 \\
		\end{array}
		\right)
	\,,\\
	&\frac{1}{b_1}\kappa =-4.621 S_{\rm g,r}^{(0)}+5.094 S_{\rm g,r}^{(1)}-0.283 S_{\rm g,r}^{(2)}
	\,,
}
for $R=10\,h^{-1}{\rm Mpc}$, 
and
\al{
	&\mathsf{M}_{\rm r}
		=\frac{1}{b_1}
        \left(
		\begin{array}{ccc}
		3.9437 & -3.2262 & 0.41878 \\
		3.8551 & -3.0347 & 0.37766 \\
		4.9700 & -4.2711 & -0.17060 \\
		\end{array}
		\right)
	\,,\\
	&\frac{1}{b_1}\kappa =-7.168 S_{\rm g,r}^{(0)}+7.789 S_{\rm g,r}^{(1)}-0.354 S_{\rm g,r}^{(2)}
	\,,
}
for $R=30\,h^{-1}{\rm Mpc}$.
We show in Fig.~\ref{fig:MInv_real-space} the coefficients of the consistency relation for the real-space skewness parameters,
namely $|[\mathsf{M}_{\rm r}^{-1}]^{(1)}{}_{(a)}|/b_1$ with $a=0$ (red solid), $a=1$ (blue dot-dashed), and $a=2$ (green dashed), as a function of the smoothing length $R\,[h^{-1}{\rm Mpc}]$.
This implies that the contribution from $S_{\rm g,r}^{(2)}$ is suppressed by an order of magnitude compared to
the other two contributions, while the other two are of the same order of magnitude.
Therefore, $S^{(0)}_{\rm g, r}$ and $S^{(1)}_{\rm g, r}$ would essentially determine the consistency relation.

Before closing this section, we would like to
comment on the scales for which
these consistency conditions are valid.
In Ref.~\cite{Matsubara:2020knr}, the authors showed that
the skewness parameters in the $N$-body simulations, which can
probe fully nonlinear regime, 
are qualitatively reproduced by the tree-level perturbation
theory on large scales within 5$\%$ precision level for
$R\gtrsim 30\,h^{-1}{\rm Mpc}$, while its accuracy decreases
to more than 10$\%$ for $R\lesssim 20\,h^{-1}{\rm Mpc}$.
Therefore, in the case of the small smoothing length, e.g., 
$R\lesssim 10\,h^{-1}{\rm Mpc}$, the nonlinear
corrections are needed in the analytic formula.

\section{Skewness consistency relation in redshift space}
\label{sec:Consistency relation in redshift space}

In redshift space, the radial position of galaxies is given by the observed radial component of its relative velocity
to an observer.
The peculiar velocity field of the underlying matter density, ${\bm v}$, distorts the observed position of the galaxy along
the line of sight.
The mapping of a galaxy from its position ${\bm x}$ in real space to its position ${\bm s}$ in redshift space
along the line-of-sight direction ${\bm n}$ is expressed as
\al{
	{\bm s}({\bm x})={\bm x}+\frac{{\bm v}\cdot{\bm n}}{aH}{\bm n}
	\,.
}
Hence, in redshift space, the density contrast should be treated as a statistically anisotropic quantity.
Even in the presence of anisotropy in the direction along the line of sight, all statistical measures should be independent
with respect to sky rotations in the plane perpendicular to the line of sight.
We will consider the skewness parameters in the plane-parallel approximation.

\subsection{Skewness parameters in redshift space}

To take into account the effect of redshift-space
distortions, we need to treat the line-of-sight 
and the perpendicular directions separately.
Hence, it is obvious that  
there are other independent parameters of skewness in anisotropic redshift space (see \cite{Codis:2013exa} for similar quantities).
In this subsection, we would like to derive the redshift-space skewness parameters describing the statistically anisotropic quantities.
In redshift space, we have several possibilities for the skewness parameters, e.g.,
\al{
	\frac{3}{2}\frac{\ave{\delta_{\rm g}|\nabla_\parallel\delta_{\rm g}|^2}_{\rm c}}{\sigma_{{\rm g}0}^2\sigma_{{\rm g}\parallel}^2}
	\,,\ \ \ 
	\frac{3}{2}\frac{\ave{\delta_{\rm g}|\nabla_\perp\delta_{\rm g}|^2}_{\rm c}}{\sigma_{{\rm g}0}^2\sigma_{{\rm g}\perp}^2}
    \,,\label{eq:S1 trial}
}
for the extension of $S^{(1)}$,
\al{
	-\frac{9}{4}\frac{\ave{|\nabla_\parallel\delta_{\rm g}|^2\Delta_\parallel\delta_{\rm g}}_{\rm c}}{\sigma_{{\rm g}\parallel}^4}
	\,,\ \ \ 
	-\frac{9}{4}\frac{\ave{|\nabla_\perp\delta_{\rm g}|^2\Delta_\perp\delta_{\rm g}}_{\rm c}}{\sigma_{{\rm g}\perp}^4}
	\,, \ \ \ 
	-\frac{9}{4}\frac{\ave{|\nabla_\perp\delta_{\rm g}|^2\Delta_\parallel\delta_{\rm g}}_{\rm c}}{\sigma_{{\rm g}\perp}^2\sigma_{{\rm g}\parallel}^2}
	\,,\ \ \ 
	-\frac{9}{4}\frac{\ave{|\nabla_\parallel\delta_{\rm g}|^2\Delta_\perp\delta_{\rm g}}_{\rm c}}{\sigma_{{\rm g}\parallel}^2\sigma_{{\rm g}\perp}^2}
	\,,\label{eq:S2 trial}
}
for the extension of $S^{(2)}$, where we have introduced
\al{
	&\sigma_{{\rm g}\parallel}^2=\ave{|\nabla_\parallel\delta_{\rm g}|^2}
	\,,\\
	&\sigma_{{\rm g}\perp}^2=\ave{|\nabla_\perp\delta_{\rm g}|^2}
	\,.
}
However, as shown below, some of these 
possibilities are not independent.
To construct the independent set of the anisotropic
skewness parameters, we first develop the {\it trial} function 
and check its behavior.
Based on this consideration, let us consider the {\it trial} functional form of the skewness parameter for the anisotropic bispectrum as
\al{
	S^{(a),{\rm trial}}_{\alpha\beta}
		=\frac{1}{\sigma_{{\rm g}0}^{4-2a}\sigma_{{\rm g}\alpha}^a\sigma_{{\rm g}\beta}^a}
		\int\frac{\dd^3{\bm k}_1}{(2\pi)^3}\frac{\dd^3{\bm k}_2}{(2\pi)^3}\frac{\dd^3{\bm k}_3}{(2\pi)^3} (2\pi)^3 \delta_{\rm D}^3({\bm k}_{123})
		 s^{(a),{\rm trial}}_{\alpha\beta} ({\bm k}_1,{\bm k}_2,{\bm k}_3)B_{\rm g,s}({\bm k}_1,{\bm k}_2,{\bm k}_3)W(k_1R)W(k_2R)W(k_3R)
	\,,\label{eq:skewness trial function}
}
where 
$B_{\rm g,s}$ denotes the galaxy
bispectrum in the redshift space, $\alpha$ represents $0,\perp, \parallel$, and we have introduced the seven kernel functions:
\al{
	&s^{(0),{\rm trial}}_{00}=1
	\,,\label{eq:s^0_00}\\
	&s^{(1),{\rm trial}}_{\parallel\parallel} =\frac{3}{4}k_{3\parallel}^2
	\,,\\	
	&s^{(1),{\rm trial}}_{\perp\perp} =\frac{3}{4}k_{3\perp}^2
	\,,\\
	&s^{(2),{\rm trial}}_{\parallel\parallel}=-\frac{9}{4}(k_{1\parallel}k_{2\parallel})k_{3\parallel}^2
	\,,\\	
	&s^{(2),{\rm trial}}_{\perp\perp}=-\frac{9}{4}({\bm k}_{1\perp}\cdot{\bm k}_{2\perp})k_{3\perp}^2
	\,,\\
	&s^{(2),{\rm trial}}_{\perp\parallel}=-\frac{9}{4}({\bm k}_{1\perp}\cdot{\bm k}_{2\perp})k_{3\parallel}^2
	\,,\\
	&s^{(2),{\rm trial}}_{\parallel\perp}=-\frac{9}{4}(k_{1\parallel}k_{2\parallel})k_{3\perp}^2
	\,.\label{eq:s^2_ab asym}
}
One can confirm that 
Eqs.~\eqref{eq:S1 trial} and \eqref{eq:S2 trial} 
can be written as
Eq.~\eqref{eq:skewness trial function} 
thanks to the symmetry of ${\bm k}_1$, ${\bm k}_2$, and ${\bm k}_3$ 
in the kernel functions of Eq.~\eqref{eq:skewness trial function}.
Hereafter, instead of using the symmetrized bispectrum, let us use the symmetrized version of $s^{(a),{\rm trial}}_{\alpha\beta}$ which is given by
\al{
	s^{(a),{\rm trial,sym}}_{\alpha\beta}({\bm k}_1,{\bm k}_2,{\bm k}_3):=\frac{1}{3}s^{(a),{\rm trial}}_{\alpha\beta}({\bm k}_1,{\bm k}_2,{\bm k}_3)+(\text{2\ cyc.})
	\,.
}
The symmetrized kernels for the spin-1 and $2$ are given by
\al{
	&s^{(1),{\rm trial,sym}}_{\alpha\alpha}
		=\frac{1}{2}\Bigl[k_{1,\alpha}^2+k_{2,\alpha}^2+{\bm k}_{1,\alpha}\cdot{\bm k}_{2,\alpha}\Bigr]
	\,,\\
	&s^{(2),{\rm trial,sym}}_{\alpha\beta}
		=\frac{3}{4}\Bigl[ k_{1\alpha}^2k_{2\beta}^2+k_{1\beta}^2k_{2\alpha}^2-2({\bm k}_{1\alpha}\cdot{\bm k}_{2\alpha})({\bm k}_{1\beta}\cdot{\bm k}_{2\beta})\Bigr]
	\,.\label{eq:s^2_ab}
}
The explicit form of the symmetrized kernel Eq.~\eqref{eq:s^2_ab} shows the symmetry of the spin-$2$ kernel function:
\al{
	&s^{(2),{\rm trial,sym}}_{\parallel\parallel}=0
	\,,\\
	&s^{(2),{\rm trial,sym}}_{\alpha\beta}=s^{(2),{\rm trial,sym}}_{\beta\alpha}
	\,,
}
which indicates
that some of the trial functions of the skewness parameters Eq.~\eqref{eq:skewness trial function} can be rewritten in terms
of other skewness parameters.
We, therefore, find that the independent number of the skewness parameters for the anisotropic bispectrum is five.
Furthermore, the kernel functions $s^{(1),{\rm trial,sym}}_{\perp\perp}$ and $s^{(2),{\rm trial,sym}}_{\perp\perp}$ can be written in terms
of the combination of $s^{(a),{\rm sym}}$ defined in Eqs.~\eqref{eq:s^1 def} and \eqref{eq:s^2 def} 
in addition to $s^{(1),{\rm trial,sym}}_{\parallel\parallel}$ and $s^{(2),{\rm trial,sym}}_{\perp\parallel}$:
\al{
	&s^{(1),{\rm trial,sym}}_{\perp\perp}
		=s^{(1),{\rm sym}}-s^{(1),{\rm trial,sym}}_{\parallel\parallel}
	\,,\\
	&s^{(2),{\rm trial,sym}}_{\perp\perp}
		=s^{(2),{\rm sym}}-2s^{(2),{\rm trial,sym}}_{\perp\parallel}
	\,.
}
Hence, we can use the real-space kernels $s^{(1),{\rm sym}}$ and $s^{(2),{\rm sym}}$ 
instead of the anisotropic kernels $s^{(1),{\rm trial,sym}}_{\perp\perp}$ and $s^{(2),{\rm trial,sym}}_{\perp\perp}$.
Furthermore, 
$s^{(1),{\rm trial,sym}}_{\parallel\parallel}$ and
$s^{(2),{\rm trial,sym}}_{\perp\parallel}$ 
include the information 
of the isotropic kernel functions such as 
$s^{(1),{\rm sym}}$
and $s^{(2),{\rm sym}}$.
To extract the independent information
of the anisotropic skewness parameters,
we also impose
that in the limit of the isotropic Universe, the anisotropic components of the skewness parameters 
reduce to zero. 

To summarize, we introduce the five independent skewness parameters in redshift space, three of which are the same as
the real-space skewness parameters defined as
\al{
	S^{(a)}_{\rm g,s}
		=\frac{1}{\sigma_{{\rm g}0}^{4-2a}\sigma_{{\rm g}1}^{2a}}
		\int\frac{\dd^3{\bm k}_1}{(2\pi)^3}\frac{\dd^3{\bm k}_2}{(2\pi)^3}\frac{\dd^3{\bm k}_3}{(2\pi)^3}(2\pi)^3\delta_{\rm D}^3({\bm k}_{123})
		 s^{(a),{\rm sym}}({\bm k}_1,{\bm k}_2,{\bm k}_3)B_{\rm g,s}({\bm k}_1,{\bm k}_2,{\bm k}_3)W(k_1R)W(k_2R)W(k_3R)
	\,,
}
with $a=0,1,2$, and remaining two characterize the anisotropy due to the redshift-space distortion, defined as 
\al{
	&S^{(3)}_{\rm g,s}
		=\frac{1}{\sigma_{{\rm g}0}^2\sigma_{{\rm g}\parallel}^2}
		\int\frac{\dd^3{\bm k}_1}{(2\pi)^3}\frac{\dd^3{\bm k}_2}{(2\pi)^3}\frac{\dd^3{\bm k}_3}{(2\pi)^3}(2\pi)^3\delta_{\rm D}^3({\bm k}_{123})
		 s^{(3),{\rm sym}}({\bm k}_1,{\bm k}_2,{\bm k}_3)B_{\rm g,s}({\bm k}_1,{\bm k}_2,{\bm k}_3)W(k_1R)W(k_2R)W(k_3R)
	\,,\label{eq:S3 def}\\
	&S^{(4)}_{\rm g,s}
		=\frac{1}{\sigma_{{\rm g}\perp}^2\sigma_{{\rm g}\parallel}^2}
		\int\frac{\dd^3{\bm k}_1}{(2\pi)^3}\frac{\dd^3{\bm k}_2}{(2\pi)^3}\frac{\dd^3{\bm k}_3}{(2\pi)^3}(2\pi)^3\delta_{\rm D}^3({\bm k}_{123})
		 s^{(4),{\rm sym}}({\bm k}_1,{\bm k}_2,{\bm k}_3)B_{\rm g,s}({\bm k}_1,{\bm k}_2,{\bm k}_3)W(k_1R)W(k_2R)W(k_3R)
	\,,\label{eq:S4 def}
}
where the convenient choice of the five independent kernel functions of the skewness parameters in redshift space is given by
the set of three known functions:
\al{
	&s^{(0),{\rm sym}}=1
	\,,\\
	&s^{(1),{\rm sym}}=\frac{1}{2}\left(k_1^2+k_2^2+{\bm k}_1\cdot{\bm k}_2\right)
	\,,\\
	&s^{(2),{}\rm sym}=\frac{3}{2}\Bigl[k_1^2k_2^2-({\bm k}_1\cdot{\bm k}_2)^2\Bigr]
	\,,
}
and two new functions:
\al{
	&s^{(3),{\rm sym}}
		=\frac{1}{2}\Bigl[k_{1\parallel}^2+k_{2\parallel}^2+k_{1\parallel}k_{2\parallel}\Bigr]-\frac{1}{3}s^{(1),{\rm sym}}
	\,,\\
	&s^{(4),{\rm sym}}
		=\frac{3}{4}
			\Bigl[k_1^2k_{2\parallel}^2+k_{1\parallel}^2k_2^2-2(k_{1\parallel}k_{2\parallel})\left({\bm k}_1\cdot{\bm k}_2\right)\Bigr]-\frac{1}{3}s^{(2),{\rm sym}}
	\,.
}

\subsection{Galaxy bispectrum in redshift space and reduced formula}

To write down the galaxy number density fluctuations
in redshift space, we need to describe the peculiar velocity field as well as the matter density field.
In the standard perturbation theory, the Fourier counterpart of
the linear velocity divergence field $\theta:=\nabla\cdot{\bm v}/aH$ can be expanded 
in a similar way as the matter density field, Eq.~\eqref{eq:delta exp}:
\al{
	\widetilde\theta({\bm k})
		=f\biggl[
			\widetilde\delta_{\rm L}({\bm k})
				+\int\frac{\dd^3{\bm k}_1}{(2\pi)^3}\frac{\dd^3{\bm k}_2}{(2\pi)^3}
					(2\pi)^3\delta_{\rm D}^3({\bm k}_{12}-{\bm k})G_2({\bm k}_1,{\bm k}_2)
					\widetilde\delta_{\rm L}({\bm k}_1)\widetilde\delta_{\rm L}({\bm k}_2)+\cdots
		\biggr]
	\,,
}
where $f$ is the logarithmic growth rate defined as $f := \dd \ln D_+ / \dd \ln a$.
Hereafter we will consider the second-order kernel $G_2$ in the form~\cite{Takushima:2013foa,Takushima:2015iha,Hirano:2018uar,Crisostomi:2019vhj,Lewandowski:2019txi,Hirano:2020dom,Yamauchi:2021nxw}:
\al{
	G_2({\bm k}_1,{\bm k}_2)=\kappa_\theta\biggl[ 1+\frac{1}{2}(\widehat{\bm k}_1\cdot\widehat{\bm k}_2)\left(\frac{k_1}{k_2}+\frac{k_2}{k_1}\right)\biggr]
			-\frac{4}{7}\lambda_\theta\biggl[1-(\widehat{\bm k}_1\cdot\widehat{\bm k}_2)^2\biggr]
	\,.\label{eq:G2}
}
Assuming the matter sector is minimally coupled to
gravity, the continuity and Euler equations for the matter give the relation between
the coefficients of $F_2$ and $G_2$ kernels as
\al{
	&\kappa_\theta =2\kappa -1+\frac{\dot\kappa}{fH}
	\,,\\
	&\lambda_\theta =\lambda +\frac{\dot\lambda}{2fH}
	\,.
}
Since the time dependence of $\kappa$ and $\lambda$ strongly depends on the underlying theory of gravity, as mentioned 
in Sec.~\ref{sec:Reduced formula for real-space galaxy bispectrum}, 
the time dependence of $\kappa_\theta$ and $\lambda_\theta$ can be also used to constrain the modification of gravity.
We note that $\kappa_\theta =1$ in the case of the Horndeski scalar-tensor theory including general relativity, but
$\kappa_\theta$ deviates from unity in the case of the beyond-Horndeski class of gravity theory~\cite{Hirano:2018uar,Crisostomi:2019vhj,Lewandowski:2019txi,Hirano:2020dom,Yamauchi:2021nxw}.

\begin{figure}
\vspace{-5mm}
\includegraphics[width=100mm]{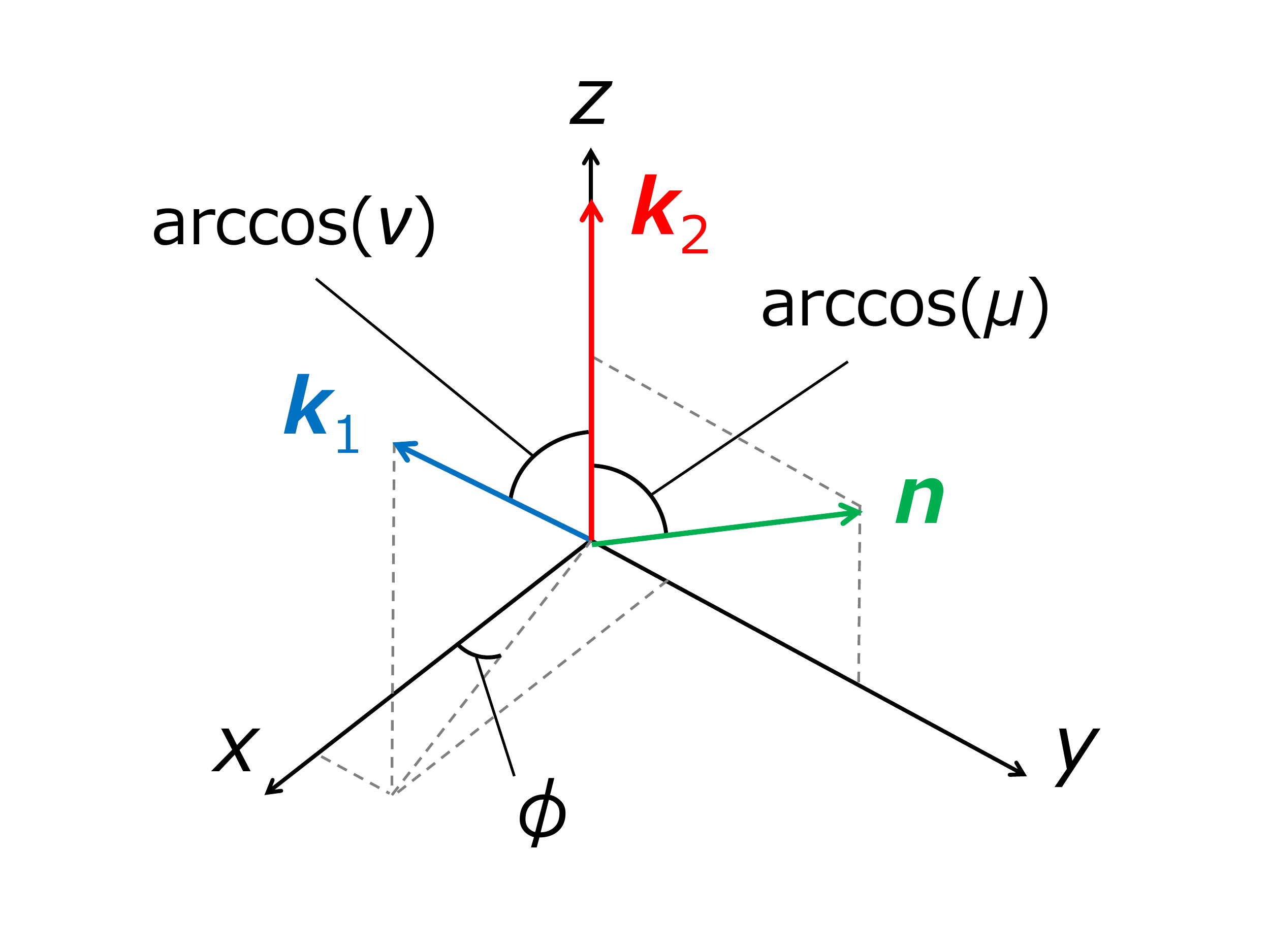}
\vspace{-5mm}
\caption{
Coordinate system and three vectors ${\bm k}_1$, ${\bm k}_2$, ${\bm n}$ for calculating the redshift-space skewness parameters.
}
\label{fig:coordinate}
\end{figure} 

The Fourier transform of the galaxy number density fluctuations in redshift space is given by
\al{
	\widetilde\delta_{\rm g,s}({\bm k})
		=Z_1({\bm k})\widetilde\delta_{\rm L}({\bm k})
			+\int\frac{\dd^3{\bm k}_1}{(2\pi)^3}\frac{\dd^3{\bm k}_2}{(2\pi)^3}(2\pi)^3\delta_{\rm D}^3({\bm k}_{12}-{\bm k})
			Z_2({\bm k}_1,{\bm k}_2)\widetilde\delta_{\rm L}({\bm k}_1)\widetilde\delta_{\rm L}({\bm k}_2)+\cdots
	\,,
}
where the linear- and second-order perturbative kernels in redshift space are \cite{Scoccimarro:1999ed}
\al{
	&Z_1({\bm k})=b_1\left(1+\beta\mu^2\right)
	\,,\label{eq:Z_1 def}\\
	&Z_2({\bm k}_1,{\bm k}_2)
		=b_1\Biggl\{	F_2({\bm k}_1,{\bm k}_2)+\beta\mu_{12}^2G_2({\bm k}_1,{\bm k}_2)
		+\frac{\beta k_{12}\mu_{12}}{2}
			\biggl[\frac{\mu_1}{k_1}
			Z_1({\bm k}_2)
			+\frac{\mu_2}{k_2}
			Z_1({\bm k}_1)
			\biggr]\Bigg\}
	\notag\\
	&\qquad\qquad\qquad
			+\frac{1}{2}b_2+b_{K^2}\biggl[(\widehat{\bm k}_1\cdot\widehat{\bm k}_2)^2-\frac{1}{3}\biggr]
	\,.\label{eq:Z_2 def}
}
Here, $\mu_a={\bm k}_a\cdot{\bm n}/|{\bm k}_a|$ with ${\bm n}$ being the line-of-sight vector, $\mu_{12}=({\bm k}_1+{\bm k}_2)\cdot{\bm n}/|{\bm k}_1+{\bm k}_2|$, and $\beta =f/b_1$.
In the tree-level approximation, the bispectrum in redshift space is given by
\al{
	B_{\rm g,s}({\bm k}_1,{\bm k}_2,{\bm k}_3)
		=2Z_1({\bm k}_1)Z_1({\bm k}_2)Z_2({\bm k}_1,{\bm k}_2)P_{\rm L}(k_1)P_{\rm L}(k_2)+(\text{cyc.})
	\,.
}
Then, the redshift-space skewness parameters can be rewritten as
\al{
	S_{\rm g,s}^{(a)}=\frac{1}{\sigma_{{\rm g}0}^{4-2a}\sigma_{{\rm g}1}^{2a}}\int\frac{\dd^3{\bm k}_1}{(2\pi)^3}\frac{\dd^3{\bm k}_2}{(2\pi)^3}
			s^{(a),{\rm sym}}({\bm k}_1,{\bm k}_2)6Z_1({\bm k}_1)Z_1({\bm k}_2)Z_2({\bm k}_1,{\bm k}_2)P_{\rm L}(k_1)P_{\rm L}(k_2)e^{-(k_1^2+k_2^2+{\bm k}_1\cdot{\bm k}_2)R^2}
	\,,
}
for $a=0,1,2$, 
and
\al{
	&S_{\rm g,s}^{(3)}=\frac{1}{\sigma_{{\rm g}0}^2\sigma_{{\rm g}\parallel}^2}\int\frac{\dd^3{\bm k}_1}{(2\pi)^3}\frac{\dd^3{\bm k}_2}{(2\pi)^3}
			s^{(3),{\rm sym}}({\bm k}_1,{\bm k}_2)6Z_1({\bm k}_1)Z_1({\bm k}_2)Z_2({\bm k}_1,{\bm k}_2)P_{\rm L}(k_1)P_{\rm L}(k_2)e^{-(k_1^2+k_2^2+{\bm k}_1\cdot{\bm k}_2)R^2}
	\,,\\
	&S_{\rm g,s}^{(4)}=\frac{1}{\sigma_{{\rm g}\perp}^2\sigma_{{\rm g}\parallel}^2}\int\frac{\dd^3{\bm k}_1}{(2\pi)^3}\frac{\dd^3{\bm k}_2}{(2\pi)^3}
			s^{(4),{\rm sym}}({\bm k}_1,{\bm k}_2)6Z_1({\bm k}_1)Z_1({\bm k}_2)Z_2({\bm k}_1,{\bm k}_2)P_{\rm L}(k_1)P_{\rm L}(k_2)e^{-(k_1^2+k_2^2+{\bm k}_1\cdot{\bm k}_2)R^2}
	\,,
}
where the variances for the anisotropic derivatives for galaxy density contrast 
are given by
\al{
	&\sigma_{{\rm g}\parallel}^2=b_1^2\int\frac{\dd^3{\bm k}}{(2\pi)^3}\left( 1+\beta\mu^2\right)^2k^2\mu^2P_{\rm L}(k)W^2(kR)
	\,,\\
	&\sigma_{{\rm g}\perp}^2=b_1^2\int\frac{\dd^3{\bm k}}{(2\pi)^3}\left( 1+\beta\mu^2\right)^2k^2\left( 1-\mu^2\right)P_{\rm L}(k)W^2(kR)
	\,.
}
These can be rewritten in term of $\sigma_{{\rm g}1}$ as
\al{
	&\sigma_{{\rm g}\parallel}^2=\left(\frac{1}{3}+\frac{2}{5}\beta+\frac{1}{7}\beta^2\right)\sigma_{{\rm g}1}^2
	\,,\\
	&\sigma_{{\rm g}\perp}^2=2\left(\frac{1}{3}+\frac{2}{15}\beta+\frac{1}{35}\beta^2\right)\sigma_{{\rm g}1}^2
	\,.
}

To evaluate the skewness parameters numerically, it is convenient to introduce the following coordinate system (see Fig.~\ref{fig:coordinate}) as
\al{
	&{\bm n}=(0,\sqrt{1-\mu^2},\mu)
	\,,\label{eq:n def}\\
	&{\bm k}_1=\frac{p}{R}(\sqrt{1-\nu^2}\cos\phi,\sqrt{1-\nu^2}\sin\phi,\nu)
	\,,\\
	&{\bm k}_2=\frac{q}{R}(0,0,1)
	\,.\label{eq:k2 def}
}
By adopting this coordinate system,
the integration can reduce to
\al{
	\int\frac{\dd^3{\bm k}_1}{(2\pi)^3}\frac{\dd^3{\bm k}_2}{(2\pi)^3}
	\to\frac{1}{32\pi^5 R^6}\int_0^\infty p^2\dd p\int_0^\infty q^2\dd q\int_{-1}^1\dd\mu\int_{-1}^1\dd\nu\int_0^{2\pi}\dd\phi
	\,,
}
and the skewness parameters can be simplified as
\al{
	&S_{\rm g,s}^{(a=0,1,2)}=\frac{b_1^3}{8\pi^4R^{2a+6}\sigma_{{\rm g}0}^{4-2a}\sigma_{{\rm g}1}^{2a}}
			\int_0^\infty\dd p\int_0^\infty\dd q
			e^{-(p^2+q^2)}\widetilde S_{\rm s}^{(a=0,1,2)}(p,q)
			P_{\rm L}\left(\frac{p}{R}\right)P_{\rm L}\left(\frac{q}{R}\right)
	\,,\label{eq:S^012 def}\\
	&S_{\rm g,s}^{(3)}=\frac{b_1^3}{8\pi^4R^{8}\sigma_{{\rm g}0}^{2}\sigma_{{\rm g}\parallel}^2}
			\int_0^\infty\dd p\int_0^\infty\dd q
			e^{-(p^2+q^2)}\widetilde S_{\rm s}^{(3)}(p,q)
			P_{\rm L}\left(\frac{p}{R}\right)P_{\rm L}\left(\frac{q}{R}\right)
	\,,\\
	&S_{\rm g,s}^{(4)}=\frac{b_1^3}{8\pi^4R^{10}\sigma_{{\rm g}\perp}^2\sigma_{{\rm g}\parallel}^2}
			\int_0^\infty\dd p\int_0^\infty\dd q
			e^{-(p^2+q^2)}\widetilde S_{\rm s}^{(4)}(p,q)
			P_{\rm L}\left(\frac{p}{R}\right)P_{\rm L}\left(\frac{q}{R}\right)
	\,,\label{eq:S^4 def}
}
where $\widetilde S_{\rm g,s}^{(A)}(p,q)$ with $A=0,\cdots, 4$ are kernel functions defined as
\al{
	&\widetilde S_{\rm g,s}^{(A)}(p,q)
		=\frac{6p^2q^2}{b_1^3}\int_{-1}^1\dd\nu e^{-pq\nu}\int_{-1}^1\frac{\dd\mu}{2} Z_1(\mu)
			\int_0^{2\pi}\frac{\dd\phi}{2\pi}\, \widetilde s^{(A),{\rm sym}}(p,q,\nu ,\mu_1)
			Z_1(\mu_1)Z_2(p,q,\mu,\nu ,\mu_1,\mu_{12})
	\,.\label{eq:tilde S^A def}
}
Here, $\widetilde s^{(0,1,2),{\rm sym}}$ in this coordinate system has the same form
introduced in Eqs.~\eqref{eq:s^0}--\eqref{eq:s^2}, and
$\widetilde s^{(3,4),{\rm sym}}$ is given by
\al{
	&\widetilde s^{(3),{\rm sym}}
		=\frac{1}{2}\Bigl[\left( p^2\mu_1^2+q^2\mu^2+pq\mu\mu_1\right)-\frac{1}{3}\left(p^2+q^2+pq\nu\right)\Bigr]
	\,,\\
	&\widetilde s^{(4),{\rm sym}}
		=\frac{3p^2q^2}{4}\Bigl[\left(\mu^2+\mu_1^2-2\mu\mu_1\nu\right)-\frac{2}{3}
		\left( 1-\nu^2\right)\Bigr]
	\,.
}

In order to evaluate the skewness parameters Eqs.~\eqref{eq:S^012 def}--\eqref{eq:S^4 def}, 
we rewrite the first- and second-order kernels in redshift space, Eqs.~\eqref{eq:Z_1 def} and \eqref{eq:Z_2 def}, in terms
of the coordinate system defined in Eqs.~\eqref{eq:n def}--\eqref{eq:k2 def}.
With these notations, the first- and second-order kernels in redshift space can be given by
\al{
	&\frac{1}{b_1}Z_1(\mu )=1+\beta\mu^2
	\,,\\
	&\frac{2}{b_1}Z_2(p,q,\mu ,\nu ,\mu_1,\mu_{12})
		=g_0\mathsf{P}_0+g_1\mathsf{P}_1+g_2\mathsf{P}_2
		+\beta\mu_{12}^2
		\Bigl( g_3\mathsf{P}_3+g_4\mathsf{P}_4\Bigr)
		+\mathsf{P}_\ast
	\,,\label{eq:Z2 redshift space}
}
where $\mathsf{P}_{0,1,2}$ was already introduced in Eq.~\eqref{eq:mathsfP real def},
and the remaining $\mathsf{P}_{3,4}$ and $\mathsf{P}_\ast$ are defined as
\al{
    &\mathsf{P}_3=\frac{1}{2}\left(\frac{p}{q}+\frac{q}{p}\right)P_1(\nu)+1
    \,,\ \ 
    \mathsf{P}_4=P_2(\nu) -1
    \,,\\
	&\mathsf{P}_\ast 
		=\beta\mu_{12}\sqrt{p^2+q^2+2pq\nu}
			\bigg\{\frac{\mu_1}{p}(1+\beta\mu^2)+\frac{\mu}{q} (1+\beta\mu_1^2)\biggr\}
	\,.\label{eq:P_ast def}
}
Here, $g_{0,1,2}$ has been defined in Eqs.~\eqref{eq:g0}--\eqref{eq:g2} and we have introduced two new variables
\al{
	&g_3=\kappa_\theta
	\,,\label{eq:g3 def}\\
	&g_4=\frac{16}{21}\lambda_\theta
	\,.\label{eq:g4 def}
}
In the limit of the Einstein-de Sitter Universe without the nonlinear bias factor, 
$g_0=34/21$, $g_1=1$, $g_2=8/21$, $g_3=1$, and $g_4=16/21$.

\subsection{Consistency relation in redshift space}

Even when the redshift-space distortion
effect is taken into account, $g_I$ ($I=0,\cdots,4$) appears in the skewness parameters linearly.
Hence the redshift-space anisotropic skewness parameters can be algebraically rewritten as
\al{
	S_{\rm g,s}^{(A)}=\sum_{I=0}^4\mathsf{M}_{\rm s}^{(A)}{}_{(I)}\,g_I+\mathsf{N}_{\rm s}^{(A)}
	\,,\label{eq:skewness relation redshift space}
}
where we have introduced the matrix $\mathsf{M}_{\rm s}$ which is defined as
\al{
	\mathsf{M}_{\rm s}^{(A)}{}_{(I)}
		:=\frac{\pd S_{\rm g,s}^{(A)}}{\pd g_I}
	\,.
}	
Unlike the real-space analysis, the above equation includes the term independent of $g_I$, $\mathsf{N}_{\rm s}$, which is directly related to the last term $P_\ast$ in Eq.~\eqref{eq:Z2 redshift space}.
The explicit forms of the components $\mathsf{M}_{\rm s}$ and $\mathsf{N}_{\rm s}$ are shown in Appendix \ref{sec:Reduced expression}.

Only when the determinant of $\mathsf{M}_{\rm s}$ is nonzero, Eq.~\eqref{eq:skewness relation redshift space} can be solved as
\al{
	g_I=\sum_{A=0}^4[\mathsf{M}_{\rm s}^{-1}]^{(I)}{}_{(A)}\left( S_{\rm g,s}^{(A)}-\mathsf{N}_{\rm s}^{(A)}\right)
	\,.\label{eq:g_I sol}
}
As we mentioned in the previous section, the nontrivial signals of gravity theories appearing in the first and third equations would be hidden by
the uncertainty of the nonlinear bias factors. Namely, the first and third equations would not be suitable for extracting the information of $g_0$ and $g_2$.
On the other hand, as for the fourth and fifth terms corresponding to the new contributions from the mapping from real space to redshift space,
Eqs.~\eqref{eq:g3 def} and \eqref{eq:g4 def} imply that there appear no bias contributions.
Hence, the redshift-space anisotropic contributions to the skewness parameters can be used to extract the information from gravity theory
without suffering from the uncertainty of the nonlinear galaxy bias.
We finally have the three meaningful consistency relations in redshift space:
\al{
	&g_1=\kappa =\sum_{A=0}^4[\mathsf{M}_{\rm s}^{-1}]^{(1)}{}_{(A)}\left( S_{\rm g,s}^{(A)}-\mathsf{N}_{\rm s}^{(A)}\right)
	\,,\label{eq:consistency redshift space 1}\\
	&g_3=\kappa_\theta =\sum_{A=0}^4[\mathsf{M}_{\rm s}^{-1}]^{(3)}{}_{(A)}\left( S_{\rm g,s}^{(A)}-\mathsf{N}_{\rm s}^{(A)}\right)
	\,,\label{eq:consistency redshift space 2}\\
	&g_4=\frac{16}{21}\lambda_\theta =\sum_{A=0}^4[\mathsf{M}_{\rm s}^{-1}]^{(4)}{}_{(A)}\left( S_{\rm g,s}^{(A)}-\mathsf{N}_{\rm s}^{(A)}\right)
	\,.\label{eq:consistency redshift space 3}
}
These are also
the main results of this paper.
When gravity is described by the Horndeski scalar-tensor theory~\cite{Horndeski:1974wa,Deffayet:2011gz,Kobayashi:2011nu} including general relativity, 
we have $\kappa =\kappa_\theta =1$, while $\lambda_\theta$ generally deviates from unity~\cite{Takushima:2013foa,Yamauchi:2017ibz}.
Therefore, Eqs.~\eqref{eq:consistency redshift space 1} and \eqref{eq:consistency redshift space 2} provide the consistency relations between
the anisotropic skewness parameters, and we can extract the information on the modification of gravity theory from Eq.~\eqref{eq:consistency redshift space 3}.
In the case of the extension of the Horndeski theory such as the DHOST theories~\cite{Langlois:2015cwa,Crisostomi:2016czh,Achour:2016rkg,BenAchour:2016fzp}, 
$\kappa$ and $\kappa_\theta$ also deviate from unity~\cite{Hirano:2018uar,Crisostomi:2019vhj,Lewandowski:2019txi,Hirano:2020dom,Yamauchi:2021nxw}.
Therefore, 
if the combinations of the skewness parameters given by the RHS of Eqs.~\eqref{eq:consistency redshift space 1} and \eqref{eq:consistency redshift space 2} depart from unity,
it would be the signals of the beyond-Honrdeski class of gravity theories.

Although the consistency relation for an arbitrary value of $\beta$ can be obtained by solving the formulae 
\eqref{eq:consistency redshift space 1}--\eqref{eq:consistency redshift space 3} numerically, in order to see the qualitative behavior of these consistency relations, we consider the small-$\beta$ approximation, which corresponds to the case of the large linear galaxy bias, such as in the case of very massive haloes.
It is expected that the asymptotes of the equations for the redshift-space skewness parameters in the limit of $\beta\to 0$ 
should reproduce the real-space ones.
The coefficient matrix $\mathsf{M}_{\rm s}$, the vector $\mathsf{N}_{\rm s}^{(A)}$, and the redshift-space skewness parameters $S_{\rm s}^{(A)}$  can be expanded in terms of $\beta$ as
\al{
	&\mathsf{M}_{\rm s}
		=\left(
		\begin{array}{c|c}
		 \mathsf{M}_{\rm r}  & 0_{2\times 3}  \\\hline
		 0_{3\times 2}  & 0_{2\times 2}  \\
		\end{array}
		\right)
		+\left(
		\begin{array}{c|c}
		\mathsf{A}  & \mathsf{B}  \\\hline
		\mathsf{C}  & \mathsf{D}  \\
		\end{array}
		\right)
		\beta +{\cal O}(\beta^2)
	\,,\label{eq:M exp}\\
	&\mathsf{N}_{\rm s}^{(A)}=\delta\mathsf{N}_{\rm s}^{(A)}\,\beta +{\cal O}(\beta^2)
	\,,\\
	&S_{\rm g,s}^{(A)}=\left(
		\begin{array}{c}
		\overline S_{\rm g,s}^{(a)}   \\\hline
		0_{1\times2}   \\
		\end{array}
		\right)
		+\delta S_{\rm g,s}^{(A)}\,
		\beta +{\cal O}(\beta^2)
	\,,
}
with $0_{m,n}$ denoting the $m\times n$ zero matrix.
The inverse of the matrix $\mathsf{M}_{\rm s}$ is approximately given as
\al{
	\mathsf{M}_{\rm s}^{-1}
		=\left(
		\begin{array}{c|c}
		0_{3\times 3}  & 0_{3\times 2}  \\\hline
		0_{2\times 3}  & \mathsf{D}^{-1}  \\
		\end{array}
		\right)\frac{1}{\beta}
		+\left(
		\begin{array}{c|c}
		\mathsf{M}_{\rm r}^{-1}  & -\mathsf{M}_{\rm r}^{-1}\mathsf{B}\mathsf{D}^{-1}  \\\hline
		-\mathsf{D}^{-1}\mathsf{C}\mathsf{M}_{\rm r}^{-1}              & \ast  \\
		\end{array}
		\right)
		+{\cal O}(\beta )
	\,,\label{eq:Minv}
}
where the symbol $\ast$ denotes
the arbitrary matrix components.
This is because this term gives only 
the ${\cal O}(\beta^2)$ subleading correction 
in the skewness consistency conditions. 
Hence, in our situation, considering 
the ${\cal O}(\beta)$ term
in $\mathsf{M}_{\rm s}^{-1}$ is enough 
to provide the leading consistency relation.
The leading order expression of \eqref{eq:g_I sol} for the first three $g_{I=0,1,2}$ 
appears at order ${\cal O}(\beta^0)$:
\al{
	&g_I=\sum_{A=0}^2[\mathsf{M}_{\rm r}^{-1}]^{(I)}{}_{(A)}S_{\rm g,s}^{(A)}+{\cal O}(\beta )\ \ \ \text{for}\ I=0,1,2
	\,.\label{eq:consistency redshift-space 1}
}
It is obvious that the above equation is 
the same as the consistency relation in real space, Eq.~\eqref{eq:consistency real space},
except for the ${\cal O}(\beta )$ correction, as expected. 
Hence, the nontrivial consistency relation in redshift space can be obtained from the remaining parts.
As for the remaining two equations for $I=3,4$, 
the ${\cal O}(\beta^0)$ equation vanishes and the next-leading order equations
at ${\cal O}(\beta)$ are given as
\al{
	&\beta g_I=\beta\biggl[
	        -\sum_{a=0}^2[\mathsf{D}^{-1}\mathsf{C}\mathsf{M}_{\rm r}^{-1}]^{(I)}{}_{(a)}\overline S_{\rm g,s}^{(a)}
			+\sum_{A=3}^4[\mathsf{D}^{-1}]^{(I)}{}_{(A)}\left(\delta S_{\rm g,s}^{(A)}-\delta\mathsf{N}_{\rm s}^{(A)}\right)
			\biggr]+{\cal O}(\beta^2 )\ \ \ \text{for}\ I=3,4
	\,.\label{eq:consistency redshift-space 2}
}
We note that one cannot use the above equation in the limit of $\beta\to 0$ since 
the inverse of the matrix $\mathsf{M}_{\rm s}$, Eq.~\eqref{eq:Minv},
in such a limit becomes singular.

Finally, let us evaluate $\mathsf{M}_{\rm s}$ and $\mathsf{N}_{\rm s}$ 
up to ${\cal O}(\beta )$ with $R=10\,h^{-1}\,{\rm Mpc}$:
\al{
	&\mathsf{M}_{\rm s}
		=\frac{1}{b_1}
        \left(
		\begin{array}{ccccc}
		 3.4930 & -2.1752 & 0.21206 & 0 & 0 \\
		 3.4049 & -1.9486 & 0.17288 & 0 & 0 \\
		 4.2484 & -3.0888 & -0.34998 & 0 & 0 \\
		 0 & 0 & 0 & 0 & 0 \\
		 0 & 0 & 0 & 0 & 0 \\
		\end{array}
		\right)
    \notag\\
		&\qquad\qquad
        +\frac{1}{b_1}\left(
		\begin{array}{ccccc}
		 2.3287 & -1.4501 & 0.14137 & 0.80182 & -1.0937 \\
		 2.2699 & -1.2991 & 0.11526 & 0.81021 & -1.0773 \\
		 2.8232 & -2.0585 & -0.23332 & 0.90152 & -1.5328 \\
		 0.86098 & -0.51106 & 0.22433  & 0.50322  & -0.57462  \\
		 0.84969 & -0.61755 & -0.07000 & 0.27046  & -0.45984  \\
		\end{array}
		\right)\beta +{\cal O}(\beta^2)
	\,,\\
	&\mathsf{N}_{\rm s}^{(A)}
		=\frac{1}{b_1}\left(
		\begin{array}{c}
		 1.6036 \\
		 1.6204 \\
		 1.8030 \\
		 0.81884 \\
		 0.54091 \\
		\end{array}
		\right)\beta
		+{\cal O}(\beta^2)
	\,,
}
which immediately leads the leading-order consistency relations \eqref{eq:consistency redshift-space 1} and \eqref{eq:consistency redshift-space 2} as
\al{
	&\frac{1}{b_1}\kappa =-4.621\overline S_{\rm g,s}^{(0)}+5.094\overline S_{\rm g,s}^{(1)}-0.283\overline S_{\rm g,s}^{(2)}+{\cal O}(\beta )
	\,,\label{eq:g1 numerical}\\
	&\beta\frac{1}{b_1}\kappa_\theta =\beta\biggl[-6.373\overline S_{\rm g,s}^{(0)}+3.085\overline S_{\rm g,s}^{(1)}+3.054\overline S_{\rm g,s}^{(2)}
			+6.051\left(\delta S_{\rm g,s}^{(3)}-\frac{0.819}{b_1}\right)-7.562\left(\delta S_{\rm g,s}^{(4)}-\frac{0.541}{b_1}\right)\biggr]+{\cal O}(\beta^2)
	\,,\label{eq:g3 numerical}\\
	&\beta\frac{1}{b_1}\frac{16}{21}\lambda_\theta =\beta\biggl[-3.748\overline S_{\rm g,s}^{(0)}+1.814\overline S_{\rm g,s}^{(1)}+2.231\overline S_{\rm g,s}^{(2)}
			+3.559\left(\delta S_{\rm g,s}^{(3)}-\frac{0.819}{b_1}\right)-6.622\left(\delta S_{\rm g,s}^{(4)}-\frac{0.541}{b_1}\right)\biggr]+{\cal O}(\beta^2)
	\,.\label{eq:g4 numerical}
}
The expression Eqs.~\eqref{eq:g1 numerical}--\eqref{eq:g4 numerical} 
are valid only when $\beta$ is nonvanishing but very small.
In the case of the realistic situation with the non-negligible value of $\beta$, the numerical evaluation
of all the components of $\mathsf{M}_{\rm s}$
and $\mathsf{N}_{\rm s}$ is needed through Eqs.~\eqref{eq:consistency redshift space 1}--\eqref{eq:consistency redshift space 3}.
An interesting observation is that the linear combination of 
the two equations \eqref{eq:g3 numerical} and \eqref{eq:g4 numerical} gives
the nontrivial equation as
\al{
    \beta\left(\frac{1}{b_1}\kappa_\theta
    -1.700\frac{1}{b_1}\frac{16}{21}\lambda_\theta\right)
        =0.739\beta\left(\overline S_{\rm g,s}^{(2)}-5\,\delta S_{\rm g,s}^{(4)}\right)
        +{\cal O}(\beta^2)
    \,,
}
which implies that 
we can extract the information of the specific 
combination of $\kappa_\theta$ and $\lambda_\theta$
from $S_{\rm g,s}^{(2)}$ and $\delta S_{\rm g,s}^{(4)}$ alone.
Although the detailed structure depends on
the definition of the skewness parameters,
this type of relationship between $\kappa_\theta$ and $\lambda_\theta$ 
can be numerically confirmed for any value of $R$.
It would be interesting to evaluate the higher-order corrections for this type of relation.

\section{Summary}
\label{sec:Summary}

In this paper, we have investigated the skewness as a possible probe to test the theory of gravity
such as the Horndeski scalar-tensor theory and the degenerate higher-order scalar-tensor (DHOST) theory.
We have developed the formula of the skewness for
the galaxy number density fluctuations
as realistic observables
and have considered how the effect of the quasi-nonlinear growth of structure appears in the skewness parameters.

We have shown that the three time-dependent parameters including the coefficients of the kernels of the second-order density contrast 
can be rewritten in terms of the linear combination of the three skewness parameters 
in the real space.
In two of the three equations, the signature of the modified gravity represented by
the second-order kernels is hidden by the uncertainty 
in the nonlinear galaxy bias functions,
but the remaining one can be used to extract the information of the gravity theory, 
in particular the DHOST theory, without suffering from the nonlinear bias 
[see Eq.~\eqref{eq:real space consistency}].
Even when the gravity is described by general relativity, the equation can be treated 
as the consistency relation 
among the three skewness parameters, which can be used as a  test of gravity theory.

We then extended the analysis from the real space to the redshift space 
and properly identified the five independent redshift-space skewness parameters 
to characterize the anisotropy due to the redshift-space distortions. 
We found that three of these are the same as the real-space skewness parameters, 
while the two new type skewness parameters are needed [Eqs.~\eqref{eq:S3 def} and \eqref{eq:S4 def}].
We have developed the analysis tool for the redshift-space skewness parameters 
and have shown the similar consistency relations to the real-space one.
Unlike the real space, we found three nontrivial consistency relations, 
two of which are related to the second-order kernels of the peculiar velocity field 
[Eqs.~\eqref{eq:consistency redshift space 1}--\eqref{eq:consistency redshift space 3}].
Using the resultant expressions, one can distinguish general relativity and 
the Horndeski scalar-tensor theory.

In this paper, we have only shown the presence of 
the consistency relations for the skewness parameters, 
but we have not addressed the forecast for
current and future large-scale observations.
Moreover, we have focused only on the specific type
of second-order kernel functions.
In the case of the Horndeski and DHOST theories
with the quasi-static approximation, these 
expressions can be shown to be valid~\cite{Hirano:2020dom}. 
However, if the gravity theory is described by other
types such as vector-tensor and bi-gravity theories,
or if the quasi-static approximation is partially broken,
there may appear correction terms in the kernel functions.
In addition, in our formulation, the dependence
on the linear galaxy bias function remains intact.
Since the bias functions in a specific type of
modified gravity theory may have the scale-dependence,
evaluating the linear galaxy bias function
under various gravity theories would be important
for the error estimation.
Hence, investigating how
the gravity theory is severely constrained
by using the consistency relation for
current and future large-scale structure
is interesting, but it is left for a future issue.

\acknowledgements

This work was supported in part by JSPS KAKENHI Grants No.~19H01891, No.~22K03627 (D.Y.), No. JP19K03835, No. 21H03403 (T.M.), No.~19K03874 (T.T.).

\appendix

\section{Reduced expression}
\label{sec:Reduced expression}

In this section, we show the explicit forms of the kernels of the skewness parameters by using the formula Eq.~\eqref{eq:tilde S^A def}.
For convenience, let us introduce the matrix and the vector defined as
\al{
	&\widetilde{\mathsf{M}}_{\rm s}{}^{(A)}{}_{(I)}(p,q)
		:=\frac{6p^2q^2}{b_1^3}\int_{-1}^1\dd\nu e^{-pq\nu}\int_{-1}^1\frac{\dd\mu}{2} Z_1(\mu)
			\int_0^{2\pi}\frac{\dd\phi}{2\pi}\, s^{(A),{\rm sym}}(p,q,\nu ,\mu_1)
			Z_1(\mu_1)\frac{\pd Z_2(p,q,\mu,\nu ,\mu_1,\mu_{12})}{\pd g_I}
	\,,\\
	&\widetilde{\mathsf{N}}_{\rm s}^{(A)}(p,q)
		:=\frac{6p^2q^2}{b_1^3}\int_{-1}^1\dd\nu e^{-pq\nu}\int_{-1}^1\frac{\dd\mu}{2} Z_1(\mu)
			\int_0^{2\pi}\frac{\dd\phi}{2\pi}\, s^{(A),{\rm sym}}(p,q,\nu ,\mu_1)
			Z_1(\mu_1)\frac{b_1}{2}\mathsf{P}_\ast (p,q,\mu,\nu ,\mu_1,\mu_{12})
	\,.
}

As a demonstration, let us consider the specific matrix components $\widetilde{\mathsf{M}}_{\rm s}^{(A=0,1,2)}{}_{(I=0,1,2)}$.
The integration of $\phi$ and $\mu$ can be done analytically 
to obtain
\al{
	\widetilde{\mathsf{M}}_{\rm s}^{(A=0,1,2)}{}_{(I=0,1,2)}
		=&\,3p^2q^2\int_{-1}^1\dd\nu
			s^{(A=0,1,2),{\rm sym}}(p,q,\nu )\mathsf{P}_{I=0,1,2}(p,q,\nu )e^{-pq\nu}
			\biggl[
				\left(1+\frac{1}{3}\beta\right)^2+\frac{4}{45}\beta^2P_2(\nu )
			\biggr]
	\,.
}
One can easily find from the above expression that the leading order terms at order ${\cal O}(\beta^0)$ coincide with
the real-space parts $\widetilde{\mathsf{M}}_{\rm r}^{(a)}{}_{(i)}:=\pd\widetilde S_{\rm r}^{(a)}/\pd g_i$ derived from Eqs.~\eqref{eq:tilde S^0}--\eqref{eq:tilde S^2}, 
and the correction terms 
appear as those with $\beta$ up to ${\cal O}(\beta^2)$.
The explicit form of the kernels after the integration of $\nu$ yields
\al{
	&\widetilde{\mathsf{M}}_{\rm s}^{(0)}{}_{(0)}
		=\left(1+\frac{1}{3}\beta\right)^2\widetilde{\mathsf{M}}_{\rm r}^{(0)}{}_{(0)}
			+\frac{8}{15}\beta^2\biggl[-3\left(\cosh (pq)-\frac{\sinh (pq)}{pq}\right) +pq\sinh (pq)\biggr]
	\,,\\
	&\widetilde{\mathsf{M}}_{\rm s}^{(0)}{}_{(1)}
		=\left(1+\frac{1}{3}\beta\right)^2\widetilde{\mathsf{M}}_{\rm r}^{(0)}{}_{(1)}
			+\frac{8}{15pq}\left(\frac{p}{q}+\frac{q}{p}\right)\beta^2
			\biggl[-\left( 9+p^2q^2\right)\left(\cosh (pq)-\frac{\sinh (pq)}{pq}\right) +3pq\sinh (pq)\biggr]
	\,,\\
	&\widetilde{\mathsf{M}}_{\rm s}^{(0)}{}_{(2)}
		=\left(1+\frac{1}{3}\beta\right)^2\widetilde{\mathsf{M}}_{\rm r}^{(0)}{}_{(2)}
			+\frac{8}{15p^2q^2}\beta^2\biggl[-6\left(9+p^2q^2\right)\left(\cosh (pq)-\frac{\sinh (pq)}{pq}\right) +\left(18+p^2q^2\right) pq\sinh (pq)\biggr]
	\,,
}
for $A=0$,
\al{
	&\widetilde{\mathsf{M}}_{\rm s}^{(1)}{}_{(0)}
		=\left(1+\frac{1}{3}\beta\right)^2\widetilde{\mathsf{M}}_{\rm r}^{(1)}{}_{(0)}
			+\frac{4}{15}\beta^2\biggl[-(3+p^2)(3+q^2)\left(\cosh (pq)-\frac{\sinh (pq)}{pq}\right) +\left( 3+p^2+q^2\right) pq\sinh (pq)\biggr]
	\,,\\
	&\widetilde{\mathsf{M}}_{\rm s}^{(1)}{}_{(1)}
		=\left(1+\frac{1}{3}\beta\right)^2\widetilde{\mathsf{M}}_{\rm r}^{(1)}{}_{(1)}
	\notag\\
	&\qquad\qquad\qquad
			+\frac{4}{15pq}\left(\frac{p}{q}+\frac{q}{p}\right)
			\beta^2\biggl[
				-\Bigl\{p^2q^2(2+p^2+q^2)+3(3+p^2)(3+q^2)+9\Bigr\}\left(\cosh (pq)-\frac{\sinh (pq)}{pq}\right)
	\notag\\
	&\qquad\qquad\qquad
				+\Bigl\{(3+p^2)(3+q^2)+3\Bigr\}pq\sinh (pq)
			\biggr]
	\,,\\
	&\widetilde{\mathsf{M}}_{\rm s}^{(1)}{}_{(2)}
		=\left(1+\frac{1}{3}\beta\right)^2\widetilde{\mathsf{M}}_{\rm r}^{(1)}{}_{(2)}
			+\frac{4}{15p^2q^2}\beta^2
				\biggl[
					-\Big\{54(5+p^2+q^2)+p^2q^2(6+p^2)(6+q^2)\Bigr\}\left(\cosh (pq)-\frac{\sinh (pq)}{pq}\right) 
	\notag\\
	&\qquad\qquad\qquad
			+\Bigl\{ 18(5+p^2+q^2)+p^2q^2(6+p^2+q^2)\Bigr\} pq\sinh (pq)
				\biggr]
	\,,
}
for $A=1$, and
\al{
	&\widetilde{\mathsf{M}}_{\rm s}^{(2)}{}_{(0)}
		=\left(1+\frac{1}{3}\beta\right)^2\widetilde{\mathsf{M}}_{\rm r}^{(2)}{}_{(0)}
			+\frac{8}{5}\beta^2\biggl[\left( 18+p^2q^2\right)\left(\cosh (pq)-\frac{\sinh (pq)}{pq}\right) -6pq\sinh (pq)\biggr]
	\,,\\
	&\widetilde{\mathsf{M}}_{\rm s}^{(2)}{}_{(1)}
		=\left(1+\frac{1}{3}\beta\right)^2\widetilde{\mathsf{M}}_{\rm r}^{(2)}{}_{(1)}
			+\frac{8}{5pq}\left(\frac{p}{q}+\frac{q}{p}\right)\beta^2
			\biggl[9\left(10+p^2q^2\right)\left(\cosh (pq)-\frac{\sinh (pq)}{pq}\right) -\left(30+p^2q^2\right)pq\sinh (pq)\biggr]
	\,,\\
	&\widetilde{\mathsf{M}}_{\rm s}^{(2)}{}_{(2)}
		=\left(1+\frac{1}{3}\beta\right)^2\widetilde{\mathsf{M}}_{\rm r}^{(2)}{}_{(2)}
    \notag\\
    &\qquad\qquad\qquad
			+\frac{8}{5p^2q^2}\beta^2\biggl[\left(810+90p^2q^2+p^4q^4\right)\left(\cosh (pq)-\frac{\sinh (pq)}{pq}\right) -6\left(45+2p^2q^2\right) pq\sinh (pq)\biggr]
	\,,
}
for $A=2$.

Hereafter, we give up showing all the results since the resultant expressions are very long.
Instead, we will show the kernels before the $\nu$ integration.
One can easily obtain the result after the $\nu$ integration analytically. 
In the case of $A=0,1,2$ with $I=3,4$, after the integration of $\phi$ and $\mu$, we have
\al{
	&\widetilde{\mathsf{M}}_{\rm s}^{(A=0,1,2)}{}_{(I=3,4)}
		=3p^2q^2\beta\int_{-1}^1\dd\nu s^{(0,1,2),{\rm sym}}(p,q,\nu )\mathsf{P}_{3,4}(p,q,\nu ) e^{-pq\nu}
	\notag\\
	&\qquad\times
			\frac{1}{105}\biggl[
				\Bigl\{ 7(5+4\beta+2\beta\nu^2)+3\beta^2(1+4\nu^2)\Bigr\}
				+\frac{4\beta(7+3\beta)pq\nu (1-\nu^2)}{p^2+q^2+2pq\nu}
			\biggr]
	\,.
}
Let us move on to the case with $A=3,4$. In this case, the integration of $\phi$ and $\mu$ leads to
\al{
	&\widetilde{\mathsf{M}}_{\rm s}^{(3)}{}_{(I)}
		=\frac{p^2q^2}{70}\int_{-1}^1\dd\nu\mathsf{P}_{I}(p,q,\nu ) e^{-pq\nu}
	\notag\\
	&\qquad\times
			\biggl[
				\left(p^2+q^2+pq\nu\right)\Bigl\{ 35+14\beta (2+\nu^2)+3\beta^2(1+4\nu^2)\Bigr\}
				+2\beta(7+3\beta)pq\nu\left(1-\nu^2\right)
			\biggr]
			-\frac{1}{3}\widetilde{\mathsf{M}}_{\rm s}^{(1)}{}_{(I)}
	\,,\\
	&\widetilde{\mathsf{M}}_{\rm s}^{(4)}{}_{(I)}
		=\frac{3p^4q^4}{70}\int_{-1}^1\dd\nu\mathsf{P}_{I}(p,q,\nu ) e^{-pq\nu}
			\Bigl\{ 35+28\beta+3\beta^2 (1+2\nu^2)\Bigr\}\left(1-\nu^2\right)
			-\frac{1}{3}\widetilde{\mathsf{M}}_{\rm s}^{(2)}{}_{(I)}
	\,,\label{eq:tilde M^4_012}
}
for $I=0,1,2$, and
\al{
	&
    \widetilde{\mathsf{M}}_{\rm s}^{(3)}{}_{(I)}
		=\frac{p^2q^2}{210}\beta\int_{-1}^1\dd\nu
			\mathsf{P}_{3,4}(p,q,\nu)e^{-pq\nu}
			\biggl[
				3pq\nu\Bigl\{63+18\beta(4+\nu^2)+5\beta^2(3+4\nu^2)\Bigr\}
	\notag\\
	&\qquad\qquad
				+\frac{1}{p^2+q^2+2pq\nu}
					\bigg\{
						21\Bigl[3(p^4+q^4)+2p^2q^2(2-5\nu^2)\Bigr]
                        +18\Bigl[(p^4+q^4)(3+2\nu^2)-2p^2q^2(-2+4\nu^2+3\nu^4)\Bigr]\beta
	\notag\\
	&\qquad\qquad\qquad\qquad\qquad
						+\Bigl[5(p^4+q^4)(1+6\nu^2)+2p^2q^2(6+3\nu^2-44\nu^4)\Bigr]\beta^2
					\biggr\}
			\biggr]
			-\frac{1}{3}\widetilde{\mathsf{M}}_{\rm s}^{(1)}{}_{(I)}
	\,,\\
	&\widetilde{\mathsf{M}}_{\rm s}^{(4)}{}_{(I)}
		=\frac{p^4q^4}{35}\beta\int_{-1}^1\dd\nu
			\mathsf{P}_{3,4}(p,q,\nu ) e^{-pq\nu}
			\biggl[
				21+9(2+\nu^2)\beta+2(1+4\nu^2)\beta^2
				+\frac{2pq\nu \beta(9+4\beta)(1-\nu^2)}{p^2+q^2+2pq\nu}
			\biggr](1-\nu^2)
	\notag\\
	&\qquad\qquad
			-\frac{1}{3}\widetilde{\mathsf{M}}_{\rm s}^{(2)}{}_{(I)}
	\,,
}
for $I=3,4$.
In the limit of $\beta\to 0$, 
one explicitly shows that 
$\widetilde{\mathsf{M}}_{\rm s}^{(3,4)}{}_{(I)}$ 
becomes zero since we imposed 
the anisotropic part of the skewness parameters in
the isotropic limit ($\beta\to 0$) reduces to zero.
We note that the ${\cal O}(\beta)$ correction term of 
the specific component,
$\widetilde{\mathsf{M}}_{\rm s}^{(4)}{}_{(0,1,2)}$,
can be rewritten in terms of
the ${\cal O}(\beta^0)$ part of
$\widetilde{\mathsf{M}}_{\rm s}^{(2)}{}_{(0,1,2)}$. 

Finally, we show the expressions for
$\widetilde{\mathsf{N}}_{\rm s}$:
\al{
	&\widetilde{\mathsf{N}}_{\rm s}^{(A=0,1,2)}
		=3p^2q^2\beta\int_{-1}^1\dd\nu s^{(0,1,2),{\rm sym}}(p,q,\nu)e^{-pq\nu}
			\Bigg[
				\frac{1}{3}\biggl\{ 2+\nu\left(\frac{p}{q}+\frac{q}{p}\right)\biggr\}
				+\frac{1}{15}\biggl\{ 10+8\nu^2+9\nu\left(\frac{p}{q}+\frac{q}{p}\right)\biggr\}\beta
	\notag\\
	&\qquad
				+\frac{1}{35}\biggl\{ 6(1+4\nu^2)+\left(11 +4\nu^2\right)\nu\left(\frac{p}{q}+\frac{q}{p}\right)\biggr\}\beta^2
				+\frac{1}{315}\biggl\{ 6+16\nu^2(3+\nu^2)+5\nu (3+4\nu^2)\left(\frac{p}{q}+\frac{q}{p}\right)\biggr\}\beta^3
			\Biggr]
	\,,
}
for $A=0,1,2$, and
\al{
	&\widetilde{\mathsf{N}}_{\rm s}^{(3)}
		=\frac{3pq}{2}\beta\int_{-1}^1\dd\nu e^{-pq\nu}
			\Biggl[
				\frac{1}{15}\Bigl\{ pq(p^2+q^2)(5+4\nu^2)+3(p^4+4p^2q^2+q^4)\nu\Bigr\}
	\notag\\
	&\qquad\qquad
				+\frac{1}{35}\Bigl\{\nu (p^4+q^4)(13+2\nu^2)+4p^2q^2\nu (11+4\nu^2)+pq(p^2+q^2)(13+32\nu^2)\Bigr\}\beta
	\notag\\
	&\qquad\qquad
				+\frac{1}{63}\Bigl\{12p^2q^2\nu(3+4\nu^2)+(p^4+q^4)\nu(13+8\nu^2)+pq(p^2+q^2)(7+48\nu^2+8\nu^4)\Bigr\}\beta^2
	\notag\\
	&\qquad\qquad
				+\frac{1}{693}\Bigl\{21(p^4+q^4)\nu (1+2\nu^2)+9pq(p^2+q^2)(1+12\nu^2+8\nu^4)+4p^2q^2\nu (15+40\nu^2+8\nu^4)\Bigr\}\beta^3
			\Biggr]
    \notag\\
    &\qquad\qquad
        -\frac{1}{3}\widetilde{\mathsf{N}}_{\rm s}^{(1)}
	\,,\\
	&\widetilde{\mathsf{N}}_{\rm s}^{(4)}
		=-\frac{9p^3q^3}{4}\beta\int_{-1}^1\dd\nu e^{-pq\nu}
			\Biggl[
				\frac{4}{15}\Big\{2pq+(p^2+q^2)\nu\Bigr\}
				+\frac{2}{35}\Bigl\{ 2pq(5+4\nu^2)+9(p^2+q^2)\nu\Bigr\}\beta
	\notag\\
	&\qquad\qquad
				+\frac{8}{315}\Bigl\{ 6pq(1+4\nu^2)+(p^2+q^2)(11+4\nu^2)\nu\Bigr\}\beta^2
	\notag\\
	&\qquad\qquad
				+\frac{2}{693}\Bigl\{5(p^2+q^2)\nu(3+4\nu^2)+2pq\left( 3+24\nu^2+8\nu^4\right)\Bigr\}\beta^3
			\Biggr](1-\nu^2)
        -\frac{1}{3}\widetilde{\mathsf{N}}_{\rm s}^{(2)}
	\,,
}
for $A=3,4$.

\end{document}